\documentclass[%
reprint,
superscriptaddress,
preprintnumbers,
 amsmath,amssymb,
 aps,
pra,
]{revtex4-2}
\usepackage{multirow}
\usepackage{subfigure}
\usepackage{graphicx}
\usepackage{dcolumn}
\usepackage{bm}
\usepackage{color}
\usepackage{dsfont}
\usepackage[normalem]{ulem}
\usepackage{hyperref}
\hypersetup{colorlinks=true,allcolors=blue}
\usepackage{wasysym}

\begin{document}
\title{Topological superconductivity in a spin-orbit coupled Kondo lattice}
\author{Zekun Zhuang}
\affiliation{Center for Materials Theory, Rutgers University, Piscataway, New Jersey 08854, USA}
\affiliation{Department of Physics, University of Wisconsin-Madison, Madison, Wisconsin 53706, USA}
\author{Piers Coleman}
\affiliation{Center for Materials Theory, Rutgers University, Piscataway, New Jersey 08854, USA}
\affiliation{Department of Physics, Royal Holloway, University of London, Egham, Surrey TW20 0EX, UK}

\date{\today }

\begin{abstract}
We consider the effect of spin-orbit coupling on a two-dimensional Kondo Lattice model, in which conduction electrons are antiferromagnetically coupled to a Yao-Lee spin liquid. When a Rashba spin-orbit interaction and a nearest-neighbor Kondo interaction is introduced, the low-energy Majorana bands become gapped and develop Chern numbers, protecting unidirectionally propagating Majorana edge modes. Our model describes a chiral topological superconductor with fractionalized charge-$e$ order parameter and spontaneously broken time-reversal symmetry, which may be of interest for certain heavy fermion superconductors, such as UTe$_2$.
\end{abstract}

\maketitle


\section{Introduction}
Though the first heavy fermion materials were discovered half a century ago\cite{menth_prl1968,ott76,steglich,ube13a}, they continue to challenge our fundamental understanding of strongly correlated quantum materials. The discovery of topological insulators\cite{Kane2005_1,Kane2005_2}, with helical boundary modes protected by nontrivial topology and time-reversal symmetry, inspired the prediction 
that analogous {\sl topological Kondo insulators}, may develop in  Kondo systems with strong spin-orbit coupling, such as SmB$_6$ \cite{Dzero2010,Alexandrov2013,Dzero2016}. On the other hand, the possible existence of topological superconductors in Kondo systems remains elusive, motivating ongoing theoretical and experimental research. 

Recently, a heavy-fermion superconductor UTe$_2$ with an extremely high upper critical field exceeding 65 T has attracted enormous attention\cite{Ran2019,Ran2019_2,Aoki2019,Aoki2022}. One of the exciting features of this material,  is that it may provide a realization of a long-sought chiral topological superconductor \cite{Hayes2021,Bae2021,Ishihara2023}, as evidenced by the observation of chiral edge states in scanning tunneling experiments \cite{Jiao2020}. The presence of chiral edge modes implies that the time-reversal symmetry (TRS) must be spontaneously broken, consistent with the early Kerr rotation experiment results \cite{Hayes2021}. However, the nature of the superconducting transition remains unclear and controversial: in improved samples, the Kerr rotation signal appears to be absent and the superconducting transition seems to be a single second-order transition, which at first sight rules out a TRS superconductor, as the underlying crystal structure allows no multidimensional irreducible Cooper pair representations \cite{Thomas2020,Thomas2021}.

In this Letter, we propose a model that describes a novel charge-$e$ topological superconductor sharing a few exotic properties with that of UTe$_2$. Firstly, it models a superconductor with spontaneous TRS breaking guaranteed by Kramer's theorem, regardless of the space group of the underlying lattice. Secondly, it is topological and possesses a chiral Majorana edge mode. While our model does not  describe UTe$_2$, the new symmetry class it represents may be of interest as a novel pairing state of non-trivial topology that lies beyond the BCS paradigm. Previous efforts to explore topological superconductivity in the Kondo lattice have relied on the Kitaev Kondo model\cite{Rosch2018}. Our model is an extension of the Kondo Lattice model considered by Coleman, Panigraphi and Tsvelik (CPT) \cite{Coleman2022,Tsvelik2022}, which describes a phase with charge-$e$ spinor order parameter and gapless Majorana excitation. Unlike the Kitaev-Kondo lattice\cite{Rosch2018}, the CPT model allows the inclusion of Kondo interactions that do not disturb the static gauge degrees of freedom associated with spin fractionalization. Our generalization of the CPT model incorporates spin-orbit coupling, a realistic feature of heavy fermion materials. We also introduce nearest-neighbor Kondo interaction, which allows for a spontaneous chiral (or sublattice) symmetry breaking and the consequential development of chiral edge states. We demonstrate that, for sufficiently large Kondo coupling and nonzero Rashba interaction, {at half-filling,} the system spontaneously breaks the global $U(1)$  and discrete TRS and chiral symmetries of the original model ($U(1)\times Z_2\times Z_2$ symmetry). The gapless Majorana band of the original CPT model  becomes gapped, acquiring a nontrivial Chern number and forming chiral Majorana edge modes. {When the chemical potential is finite, our model describes a TRS-breaking charge-$e$ topological superconductor, the transition of which from the normal state is of single-step.}

\section{Model}\begin{figure}[t]
    \centering
    \includegraphics[width=0.48\textwidth]{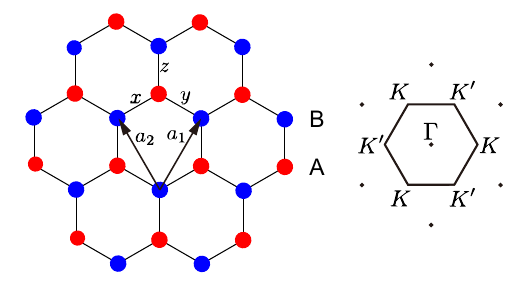}
    \caption{Left: The honeycomb lattice where $x,y,z$ denotes the type of bonds. Right: The reciprocal lattice and first Brillouin zone of the honeycomb lattice.}
    \label{fig:lattice}
\end{figure}
We consider coupling a Yao-Lee (YL) spin liquid to a conduction sea on the honeycomb lattice described by the Hamiltonian $H=H_C+H_{YL}+H_K$, where
\begin{eqnarray}
    H_C&=&\sum_{\langle ij\rangle} c_{i}^\dagger \left[-tI+i\lambda_R(\Vec{\sigma}\times \Vec{d}_{ij}) \cdot \hat{z}\right]c_{j }+\text{H.c.}, \label{FermiSeaH}\\
     H_{YL}&=&\frac{K}{2}\sum_{\langle ij\rangle}(\Vec{\sigma}_i\cdot\Vec{\sigma}_j)\lambda^{\alpha_{ij}}_i\lambda^{\alpha_{ij}}_j,\label{YLH}
    \\
    H_{K}&=&\frac{1}{2}\sum_{i,j} J_{ij}\Vec{\sigma}_i\cdot(c_{j}^\dagger \Vec{\sigma} c_{j }).
\end{eqnarray}
Here $H_C$ describes the conduction electron Hamiltonian in the presence of Rashba spin-orbit interaction
where $I$ is the identity matrix, $\vec{\sigma}$ is the Pauli matrix associated with spins, $\langle ij\rangle$ denotes the pair of nearest-neighbor (NN) sites $(i,j)$ where $i\in A, j\in B$,  $\Vec{d}_{ij}$ represents the vector pointing from site $j$ to $i$, and we have used the shorthand $c_j^\dagger\equiv (c_{j\uparrow}^\dagger,c_{j\downarrow}^\dagger)$. $H_{YL}$ describes the Yao-Lee spin liquid\cite{Yao2011},
in which $\vec \sigma_i$ and $\vec \lambda_i$ are Pauli matrices denoting the spin and orbital degrees of freedom at each site $i$, 
while $\alpha_{ij}=x,y,z$ denotes the direction of the Ising coupling between neighboring orbitals (see Fig. \ref{fig:lattice}). $H_{K}$ describes the  Kondo coupling between the conduction electrons and local moments, where  $J_{ij}=J_1\delta_{ij} + J_2 \eta_{\langle ij\rangle}$ where $\eta_{\langle ij\rangle }=1 $ when $i$ and $j$ are nearest neighbors and $J_{1,2}\geq 0$  are antiferromagnetic. 

It is convenient to make a global gauge transformation for conduction electrons on the $A$ sublattice, $(c_{A}^\dagger,c_{B}^\dagger)\rightarrow (i c_{A}^\dagger,c_{B}^\dagger)$, so  $H_C$ becomes
\begin{equation}
    \tilde{H}_C=-\sum_{\langle ij\rangle} c_{i}^\dagger \left[itI+\lambda_R(\Vec{\sigma}\times \Vec{d}_{ij}) \cdot \hat{z}\right]c_{j }+\text{H.c.}. \label{FermiSeaH2}
\end{equation}

The YL model can be exactly solved using the Majorana representation $\Vec{\sigma}_j=-i\Vec{\chi}_j\times \Vec{\chi}_j$ and $\Vec{\lambda}_j=-i\Vec{b}_j\times \Vec{b}_j$ with the constraint $D_j=8i\chi_j^1\chi_j^2\chi_j^3 b_j^1b_j^2b_j^3=1$ \cite{Coleman1993,Kitaev2006,Yao2011}. Here the Majorana fermions $\chi, b$ obey anticommutation relations $\{\chi_i^\alpha,\chi_j^\beta\}=\{b_i^\alpha,b_j^\beta\}=\delta_{ij}\delta^{\alpha \beta}$ and $\{\chi_i^\alpha,b_j^\beta\}=0$. Equation (\ref{YLH}) then becomes 
\begin{equation}
    \tilde{H}_{YL}=iK\sum_{\langle ij\rangle} u_{ij}(\Vec{\chi}_i\cdot \Vec{\chi}_j), \label{YLMajoranaH}
\end{equation}
where $u_{ij}=-2i b_i^{\alpha_{ij}} b_j^{\alpha_{ij}}=\pm 1$ are independent, static $Z_2$ gauge fields that commute  with the Hamiltonian. Note that unlike a Kitaev Kondo lattice, where the spin-operator creates visons, the Yao-Lee structure of the CPT model means that the spin operator $\vec S_j$ commutes with the gauge fields so that action of the Kondo interaction does not disturb the static gauge fields. The ground state of Eq. (\ref{YLMajoranaH}) lies in the fluxless gauge sector \cite{Lieb1994}, where the gauge fluxes $W_p=\prod_{i,j \in p} u_{ij}=1$ (here we use the convention $i\in A, j\in B$). Provided the flux gap is larger than all other relevant energy scales, we can fix the ground-state gauge to be $u_{ij}=1$ whereupon  Eq. (\ref{YLMajoranaH}) becomes
\begin{equation}
    H_{YL}=iK\sum_{\langle ij\rangle} (\Vec{\chi}_i\cdot \Vec{\chi}_j). \label{YLMajoranaH2}
\end{equation}
In terms of Majorana fermions $\vec{\chi}$, the Kondo interaction $H_K$ becomes
 \begin{equation}
     H_{K}=-\sum_{i,j}\frac{J_{ij}}{2}(c_j^\dagger \Vec{\sigma} \cdot \Vec{\chi}_i)(\Vec{\chi}_i\cdot \Vec{\sigma} c_j).\label{HK}
 \end{equation}

To construct a mean-field solution we need to understand the underlying symmetries.  The  Hamiltonian $\tilde{H}=\tilde H_C+ H_{YL}+H_K$ (\ref{FermiSeaH2},\ref{YLMajoranaH2},\ref{HK}) has three discrete symmetries:

(1) Time-reversal $\mathcal{\tilde T}=\mathcal{TG}$, where $\mathcal{T}c_j\mathcal{T}^{-1}=i\sigma_y c_j$, $\mathcal{T}i\mathcal{T}^{-1}=-i$, $\mathcal{T}\vec{\chi}_j\mathcal{T}^{-1}=\vec{\chi}_j$ and $\mathcal{G}$ is the global $Z_2$ gauge transformation that flips the signs on the B sublattice: $\mathcal{G} c_{A(B)} \mathcal{G}^{-1}=\pm c_{A(B)}$, $\mathcal{G} \vec{\chi}_{A(B)} \mathcal{G}^{-1}=\pm \vec{\chi}_{A(B)}$.

(2) Spin-lattice rotation $\mathcal{C}=\mathcal{RUG}$, where $\mathcal{R}$ and $\mathcal{U}$ rotate the lattice and spin clockwise by $\pi/3$: $\mathcal{U} c_j \mathcal{U}^{-1}=\exp ({i\frac{\pi}{6}\sigma^z}) c_j$, $\mathcal{U} \vec{\chi}_j \mathcal{U}^{-1}=R^{-1}_z(\frac{\pi}{3})\vec{\chi}_j$, where $R_z(\phi)$ is the $SO(3)$ matrix that rotates a vector around $z$-axis by $\phi$; $\mathcal{R} c_j\mathcal{R}^{-1}=c_{j^\prime}$, $\mathcal{R} \vec{\chi}_j\mathcal{R}^{-1}=\vec{\chi}_{j^\prime}$, where $\mathbf{r}_{j^\prime}=R_z(\frac{\pi}{3})\mathbf{r}_{j}$ and $\mathbf{r}_{j}$ is the position vector of site $j$.  The additional  $Z_2$ gauge transformation ${\mathcal G}$ in $\mathcal{C}$ is needed to preserve the rotational invariance of the hopping terms in the Hamiltonian. 

(3) Sublattice particle-hole, or ``chiral'' symmetry:  $\mathcal{S}$: $\mathcal{S} i\mathcal{S}^{-1}=-i$, $\mathcal{S} c_{A(B)}\mathcal{S}^{-1}=\pm c^\dagger_{A(B)}$, and $\mathcal{S} \vec{\chi}_{A(B)} \mathcal{S}^{-1}=\pm \vec{\chi}_{A(B)}$.

\section{Mean-field solution}
With a Hubbard-Stratonovich transformation, the local and nearest-neighbor Kondo interactions become \cite{Coleman2022,Tsvelik2022}
\begin{equation}
    H_{K1}=\sum_j\left\{[(c_j^\dagger \Vec{\sigma} V_j)\cdot\Vec{\chi}_j +\text{H.c.}]+2\frac{|V_j|^2}{J_1}\right\}, \label{HK1MF}
\end{equation}
\begin{equation}\label{HK2MF}
    H_{K2}=\sum_{\langle ij\rangle}\left\{ [(c_i^\dagger (\Vec{\sigma}\cdot \Vec{\chi}_j) V_{ji})+\text{H.c.}] + 2\frac{| V_{ij}|^2}{J_2} + (i\leftrightarrow j)\right\}, 
\end{equation}
where $V_j$ and $V_{ij}$ are charge-$e$ spinors: stationarity with respect to variations in these spinors imposes the self-consistency conditions:
\begin{equation}
    V_j=-\frac{J_1}{2}\langle (\Vec{\chi}_j\cdot \Vec{\sigma}) c_j\rangle,\quad V_{ij}=-\frac{J_2}{2}\langle (\Vec{\chi}_i\cdot \Vec{\sigma} )c_j\rangle.\label{SCcondition1}
\end{equation}

If $J_2=0$,  beyond a critical value of $J_1$, a uniform spinor condensate  develops, in which   $V_j=W_1v_j/\sqrt{2}$, where $W_1>0$ and we may take $v_j=e^{i\alpha}(1,0)^T$ \cite{Tsvelik2022}. \begin{figure}[t]
    \centering
    \includegraphics[width=0.5\textwidth]{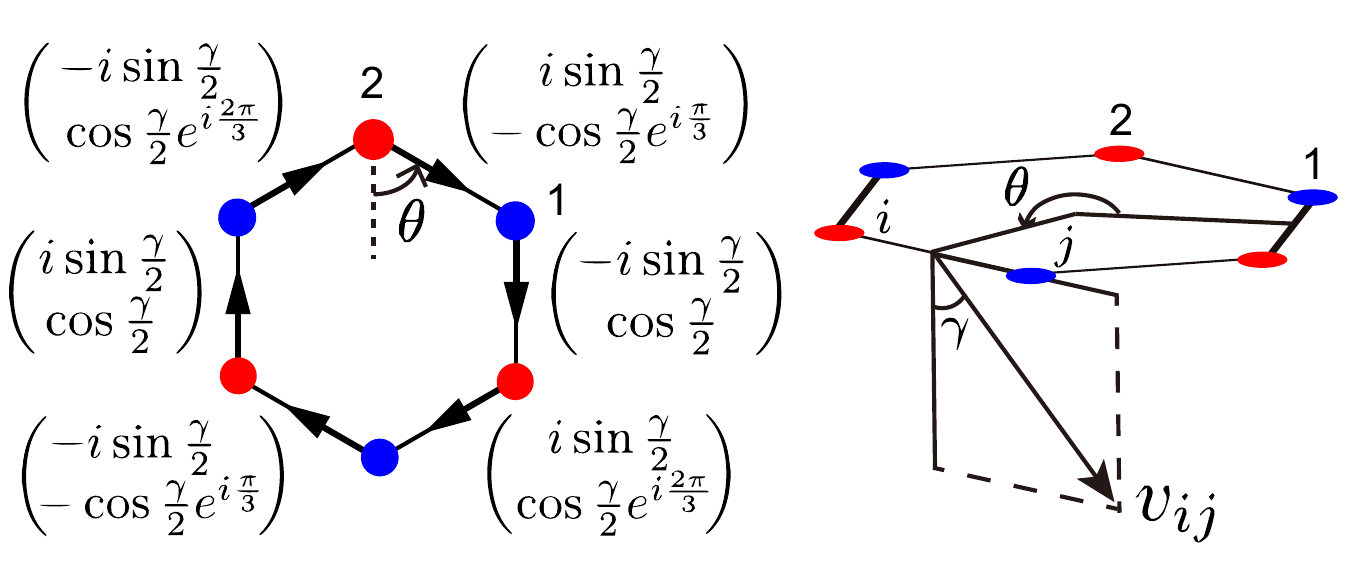}
    \caption{{Left: the mean-field ansatz of the normalized spinor order parameter $v_{ij}$, denoted by the arrow pointing from site $j$ to site $i$; Right: the direction of the spinor. We have taken $\alpha=\beta=0$ and $\phi=-\pi/2$ as stated in the text.}}
    \label{fig:ansatz}
\end{figure}
When $J_2>0$, a second spinor order parameter $V_{ij}=W_2v_{ij}/\sqrt{2}$ ($W_2>0$) also develops at sufficiently large $J_2$. As we will show below, when $W_1$ or $W_2$ becomes nonzero, the model describes a superconductor, which has superfluid stiffness proportional to $W_1^2, W_2^2$. {The ansatz for $v_{ij}$ is parametrized by
\begin{align}\label{ans}
        v_{ij}&=e^{i(\alpha+\beta)}\left(\begin{array}{c}
         -\sin \frac{\gamma}{2}e^{-i\phi}e^{3i\theta_{ij}} \\
         \cos \frac{\gamma}{2}e^{4i\theta_{ij}}
    \end{array}\right),
\end{align}
which is the spin-down spinor along the axis $\hat{n}=(\sin \gamma \cos (\phi+\theta_{ij}),\sin \gamma\sin(\phi+\theta_{ij}),\cos \gamma)$.} Here $\theta_{ij}$ is the angle between $\vec{d}_{ij}$ and $-\hat{y}$, as shown in Fig. \ref{fig:ansatz}. The ansatz (\ref{ans}) has been chosen so that both $v_i$ and $v_{ij}$ transform under the same spin $\frac{1}{2}$ representation under a $60^\circ$ spin-lattice rotation:  $v_i\rightarrow e^{i\pi/6}v_i$, $v_{ij}\rightarrow e^{i\pi/6}v_{ij}$.

The low-energy degrees of freedom of the system involve fluctuations in the overall phase $\alpha$,  which couple to the external electromagnetic field. Under a spin-lattice rotation $\mathcal{C}$, $e^{i\alpha}\rightarrow e^{i(\alpha+\pi/6)}$ while the relative $U(1)$ phase $\beta$ is unchanged, indicating that $v_j$ and $v_{ij}$ transform under the same one-dimensional irreducible representation  (see Appendix \ref{App:A} for a detailed discussion of the symmetry transformation of the order parameters). We have verified that the energy is minimized for $\beta=0,\pi$. This residual $Z_2$ degeneracy originates from spontaenous chiral symmetry breaking: under the chiral operation {$\mathcal{S}$, $v_j\rightarrow -v_j$ while $v_{ij}\rightarrow v_{ij}$}. It is this  broken chiral symmetry that is responsible for chiral edge states, demonstrated below. {The other two degrees of freedom $\phi$ and $\gamma$ are generally gapped and hence do not affect the low-energy physics.} Kramers' theorem implies there are two other states which are the time-reversal partners of the above $Z_2$ states, with order parameters given by $v_j\rightarrow-i\sigma_y  v_j^*$, $v_{ij}\rightarrow i\sigma_y  v_{ij}^*$, leading to  4-fold degenerate $U(1)\times Z_2\times Z_2$ state, where time-reversal, chiral, and electromagnetic $U(1)$ gauge symmetry are spontaneously broken. From now we only focus on the representative case $v_j=(1,0)^T$ by gauge-fixing $\alpha=0$ and absorbing the phase factor $e^{i\beta}=\pm 1$ into  $W_2$ by allowing it to acquire both positive and negative values. {In fact, we find the energy is minimized only when $\phi=\pm\pi/2$, allowing us to fix $\phi=-\pi/2$ and $\gamma\in \left[-\pi,\pi\right]$}. 
Without loss of generality, we can take $\lambda_R$ in \eqref{FermiSeaH2} to be positive, noting that  {the opposite case is obtained by a global spin rotation through 180$^\circ$ around the $z$-axis, under which $\lambda_R\rightarrow -\lambda_R$, $\gamma\rightarrow -\gamma$ and $W_2\rightarrow -W_2$}. 

 It is useful\cite{Tsvelik2022} to decompose the conduction electrons into a scalar and vector Majorana components, $c^0$ and  $\vec{c}$ respectively,
\begin{equation}
        \left(\begin{array}{c}
         c_{j,\uparrow} \\
          c_{j,\downarrow}
    \end{array}\right)=\frac{1}{\sqrt{2}}\left(\begin{array}{c}
         c_j^0+ic_j^3 \\
          ic_j^1-c_j^2
    \end{array}\right). \label{MajoranaSub}
\end{equation}

The key feature that then emerges, is that when $W_2=\lambda_R=0$, the vector components $c^{\alpha}_j$ ($\alpha =1,2,3$) develop a gap by selectively hybridizing with the Yao-Lee spin liquid, while the scalar components $c^0_j$ decouple,  forming a gapless Majorana sea. The corresponding Hamiltonian is
\begin{align}
    H=&\sum_{\mathbf{k}\in \frac{1}{2}\text{BZ}}\sum_{\alpha=1}^3\left(\begin{array}{cc}
        c_{\mathbf{k}}^{\alpha\dagger} & \chi_{\mathbf{k}}^{\alpha\dagger}
    \end{array}\right)\left(\begin{array}{cc}
        -t \vec{A}_\mathbf{k}\cdot \vec{\tau} &  -iW_1\\
        iW_1 & K \vec{A}_\mathbf{k}\cdot \vec{\tau}
    \end{array}\right)\left(\begin{array}{c}
        c_{\mathbf{k}}^{\alpha}\\
        \chi_{\mathbf{k}}^{\alpha}
    \end{array}\right)\\ \nonumber
    &+\sum_{\mathbf{k}\in \frac{1}{2}\text{BZ}} c_{\mathbf{k}}^{0\dagger}\left(-t\vec{A}_\mathbf{k}\cdot \vec{\tau}\right)c_{\mathbf{k}}^{0},
\end{align}
where $\vec{\tau}$ are the Pauli matrices spanning the sublattice degrees of freedom and  the form-factor $A_\mathbf{k}=1+e^{-i\mathbf{k}\cdot \mathbf{a_1}}+e^{-i\mathbf{k}\cdot \mathbf{a_2}}$ (where $\mathbf{a_{1,2}}=(\pm\sqrt{3}/2,3/2)$ are the Bravais lattice vectors) has been rewritten as a vector $\vec{A}_\mathbf{k}=-(\text{Im} A_\mathbf{k}, \text{Re} A_\mathbf{k},0)$. Here we have defined Fourier transforms of Majorana fermions $c^\alpha$, $\vec{\chi}$ 
\begin{align}
        c_{\mathbf{k}}^{\alpha\dagger}&=\frac{1}{\sqrt{N}}\sum_j e^{i\mathbf{k}\cdot \mathbf{R}_j}\left(c_{jA}^\alpha, c_{jB}^\alpha\right),\\\nonumber
        \vec{\chi}_{\mathbf{k}}^{\dagger}&=\frac{1}{\sqrt{N}}\sum_j e^{i\mathbf{k}\cdot \mathbf{R}_j}\left(\vec{\chi}_{jA}, \vec{\chi}_{jB}\right),
\end{align}
where $\mathbf{R}_j$ is the position of the unit cell. The summations run over half of the Brillouin zone (BZ),  due to the redundancies $\vec\chi_{\mathbf{k}}^{\dagger}=\vec\chi_{\mathbf{-k}}$ and $c_{\mathbf{k}}^{\alpha\dagger}=c_{\mathbf{-k}}^{\alpha}$.

\begin{figure}[b]
    \centering
    \includegraphics[width=0.4\textwidth]{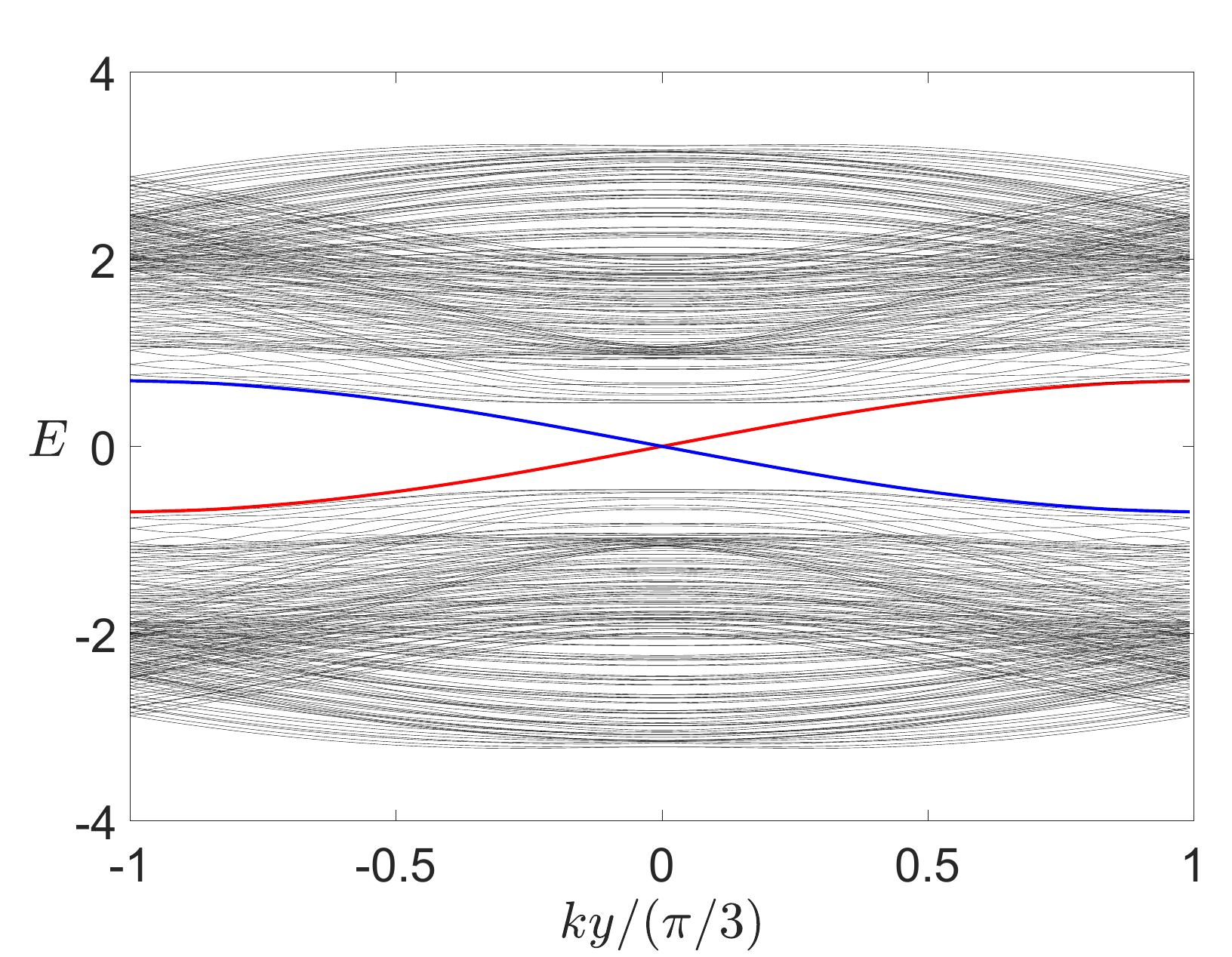}
    \caption{Spectrum obtained by exact diagonalization of the full Hamiltonian in a ribbon geometry, which is infinite in the $y$-direction and has $N_x$ unit cells in the $x$-direction. The black lines denote the bulk modes while the red and blue lines represent the unidirectional boundary modes on opposite edges. Parameters: $t=K=W_1=1$, $W_2=0.3$, $\lambda_R=0.3$, {$\gamma=0$, $\mu=0$}, $N_x=20$. }
    \label{fig:edge}
\end{figure}
The energies of the  gapped vector fermions are $\pm E_\mathbf{k}^{+}$ and $ \pm E_\mathbf{k}^{-}$, where 
\begin{equation}
    E_\mathbf{k}^{\pm}=\sqrt{\left(\frac{K+t}{2}|{A}_\mathbf{k}|\right)^2+W_1^2}\pm \left(\frac{K-t}{2}\right)|{A}_\mathbf{k}|,
\end{equation}
while the energy of the gapless scalars, $c^0$ is 
 $E_\mathbf{k}^{0}=\pm t|{A}_\mathbf{k}|$.

When $\lambda_R$ and $W_2$ are nonzero, matrix elements develop which link the low-energy subspace $|\mathcal{L}\rangle$ of gapless scalars  $c^0$ to the high-energy subspace $|\mathcal{H}\rangle$ of gapped vector Majoranas (See Appendix \ref{App:B} for the full mean-field Hamiltonian). Using second-order perturbation theory (see Appendix \ref{App:C} for detailed calculation), the low-energy effective Hamiltonian near the high-symmetry momenta $K(K^\prime)=(\pm 4\pi/3\sqrt{3},0)$ is given by
\begin{equation}
    H_\text{eff}(k)= c_{\mathbf{k}}^{0\dagger}\left[ v_F \left(\pm k_x\tau_y- k_y\tau_x\right)\mp m\tau_z)\right]c_{\mathbf{k}}^{0}, \label{effH}
\end{equation}
where $v_F=3t/2$ is the Fermi velocity, {$m=9W_2\lambda_R \cos (\gamma/2)/W_1$} is the mass and we have taken $t=K$ for simplicity. It is important to note that the mass has an opposite sign at different valleys $K,K^\prime$, indicating that the low-energy band is topological with a non-trivial Chern number $\mathcal{C}=\text{sgn}(m)$, resembling the Kitaev and Haldane models \cite{Kitaev2006,Haldane1988}. 

\begin{figure}[tb]
    \centering
    \includegraphics[width=0.4\textwidth]{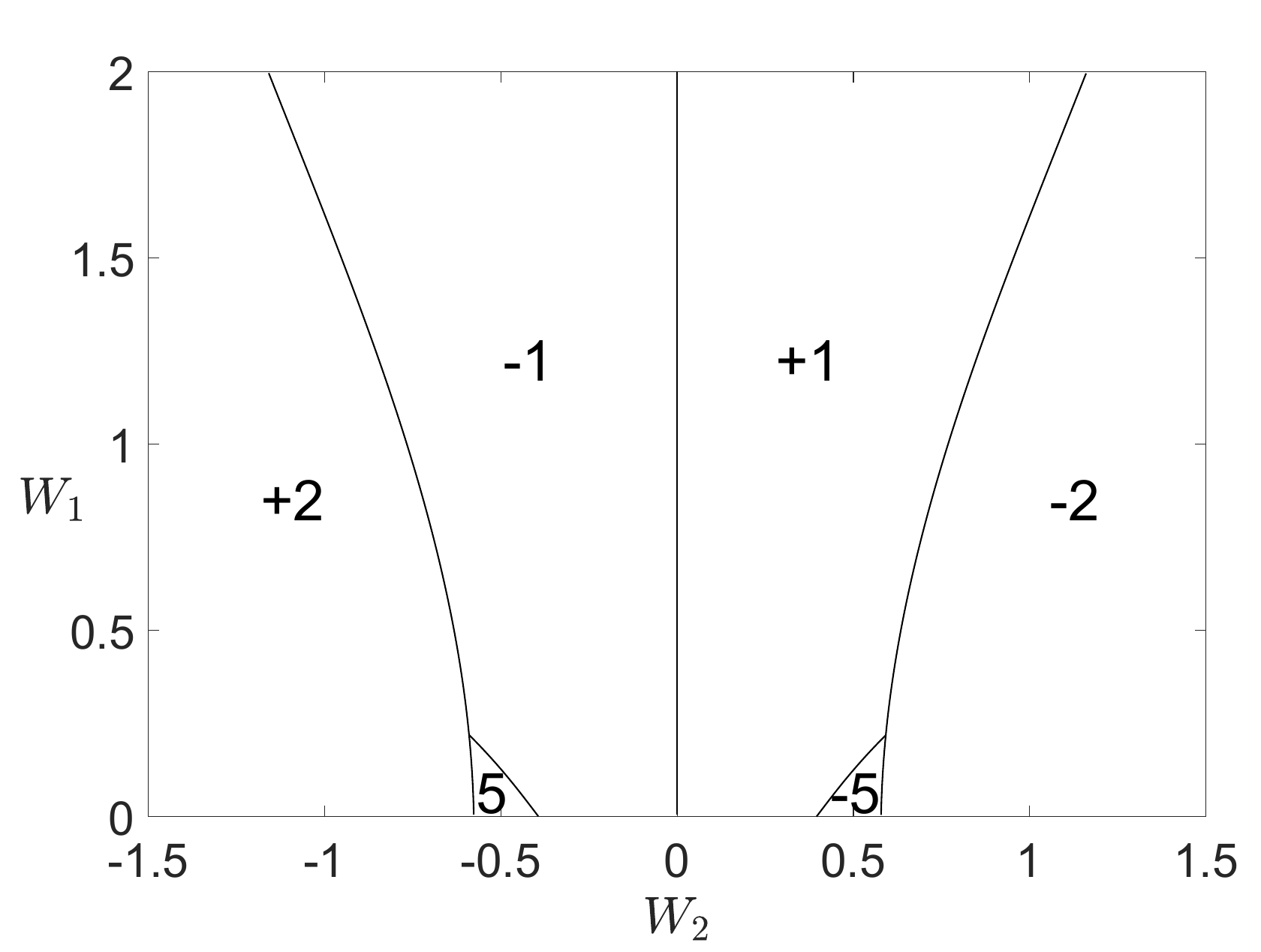}
    \caption{Topological phase diagram as a function of $W_1$ and $W_2$,   which are proportional to the amplitude of the spinor order parameters defined in Eq. (\ref{SCcondition1}). We choose $t=K=1$, $\lambda_R=0.3$, $\gamma=0$, $\mu=0$. The numbers indicate the Chern number of the corresponding topological phase.}
    \label{fig:W1W2diagram}
\end{figure}

As the high-energy bands are topologically trivial and the order parameter carries $U(1)$ electric charge, we conclude that the system is a charge-$e$ topological superconductor, which, at the mean-field level, belongs to the symmetry class D in the Altland-Zirnbauer classification \cite{Altland1997,Ludwig2016}. Due to the bulk-edge correspondence and the fact that the low-energy gapped excitations are Majorana fermions (or complex fermions living in half of the Brillouin zone), there must be a chiral  Majorana edge mode, as shown in Fig. \ref{fig:edge}. For larger $W_2$ and $\lambda_R$ where the perturbative analysis breaks down, the bulk gap may close and reopen, leading to other topological phases with different Chern numbers. In Fig. \ref{fig:W1W2diagram}, we show the $W_1$-$W_2$ phase diagram {for $\gamma=0$} obtained by exact diagonalization of the mean-field Hamiltonian, from which one finds phases with other higher Chern numbers {(see Appendix \ref{App:D} for phase diagrams at other values of $\gamma$)}. We note that the phase diagram is symmetric about $W_2=0$ and the Chern number changes sign if $W_2$ is opposite. This is a consequence of the chiral symmetry, recalling that the chiral operation reverses the chirality (velocity) of the edge modes as well as the sign of $W_2$.

\begin{figure*}
    \subfigure[]{\label{fig:phasediagram1}
    \includegraphics[width=0.3 \textwidth]{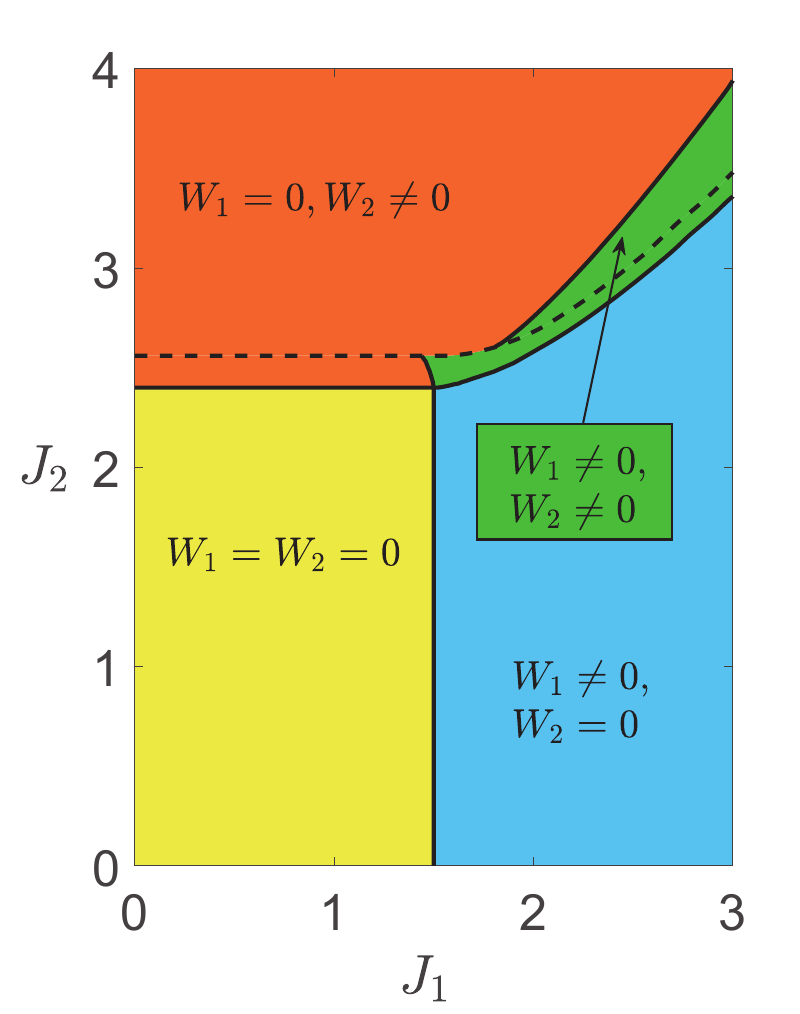}}
    \subfigure[]{\label{fig:phasediagram2}
    \includegraphics[width=0.3 \textwidth]{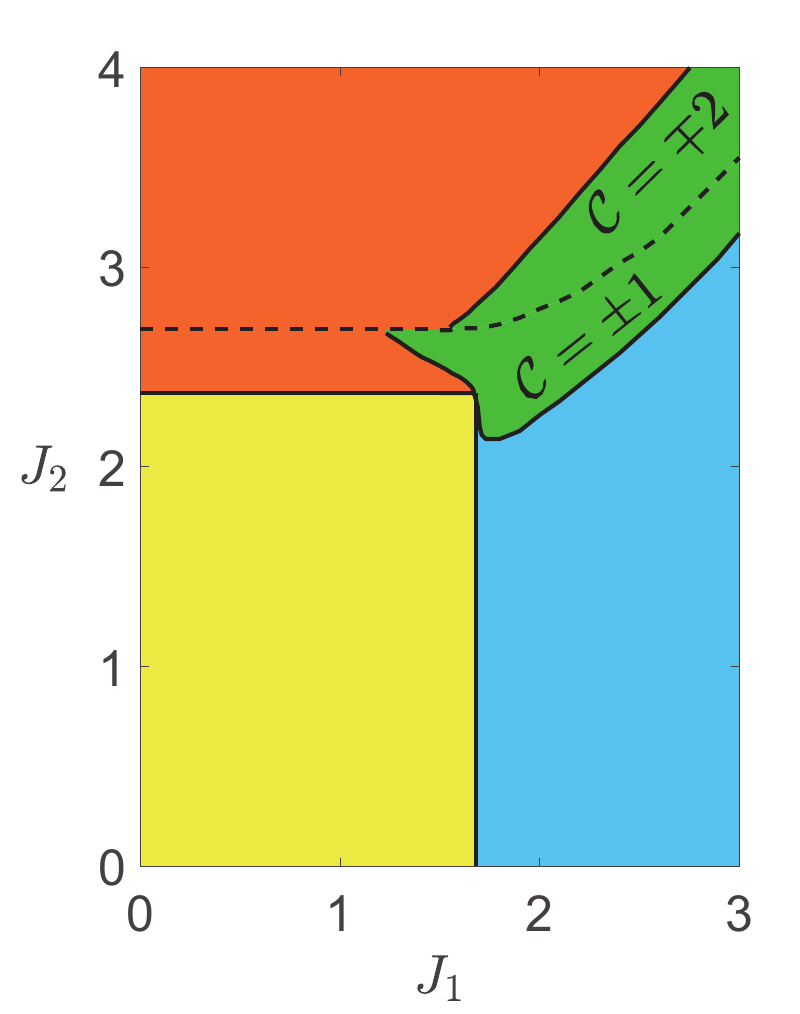}}  
        \subfigure[]{\label{fig:phasediagram3}
    \includegraphics[width=0.3 \textwidth]{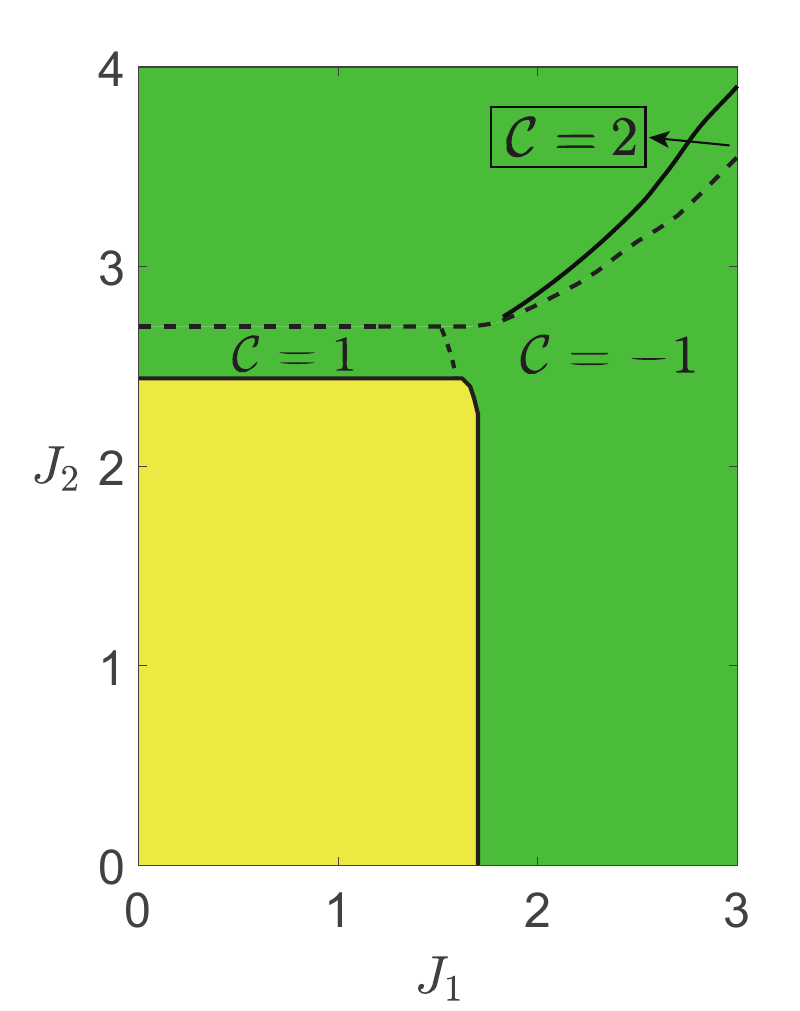}}  
    \caption{Mean-field phase diagram for (a) $\lambda_R=0, \mu=0$ (b) $\lambda_R=0.3, \mu=0$ (c) $\lambda_R=0.3, \mu=0.1$ when $t=K=1$. The solid (dashed) line denotes the continuous(discontinuous) phase transition. Here $J_1$ and $J_2$ are the on-site and nearest-neighbor Kondo coupling respectively.}
   \label{fig:phasediagram} 
\end{figure*}

The $J_1-J_2$ phase diagrams are obtained by solving the self-consistent mean-field equations numerically (see Appendix \ref{App:B} for the expression of mean-field equations) and depicted in Fig. \ref{fig:phasediagram}(a)(b). For sufficiently large $J_1$ and $J_2$, a phase region with nonzero $W_1$ and $W_2$ exists. Remarkably we find both the $\mathcal{C}=\pm 1$,  $\mathcal{C}=\mp 2$ topological superconductor phases, which are separated by a first-order transition with a sudden jump of order parameters.

{It is essential to note that the second-order transition from the disordered phase to the topological phase is generally two-step. This can be understood by writing down the Landau theory for the order parameters $W_1,W_2$: $\mathcal{L}=c_1 W_1^2+c_2W_2^2+...$, where the $W_1W_2$ term is absent due to the chiral symmetry. As $W_1$ and $W_2$ are not coupled to the lowest order, one expects that their condensations are generally independent. To make a single-step second-order transition possible, we introduce a chemical potential term 
\begin{equation}
    H_\mu=-\mu \sum_i c_i^\dagger c_i=i\mu\sum_i (c_i^3 c_i^0+c_i^2 c_i^1),
\end{equation}
which explicitly breaks the chiral symmetry for $\mu\neq 0$ and couples $W_1$ and $W_2$. Consequently, when $\mu\neq 0$, $W_1$ and $W_2$ condense simultaneously for sufficiently large Kondo coupling, and the 4-fold degenerate state becomes the 2-fold degenerate $U(1)\times Z_2$ state, as shown in Fig. \ref{fig:phasediagram}(c). Remarkably, the topological phases we found earlier are robust against the nonzero chemical potential $\mu$, unless it closes and reopens the gap. Therefore, when $\mu\neq 0$, the TRS-breaking transition to a (possibly topological) superconductor becomes one-step.
}
\section{Conclusion}

We have presented calculations on the CPT Kondo lattice which demonstrate the feasibility of topological superconductivity in the Kondo lattice.  Our approach takes advantage of an underlying Yao-Lee spin model to control the gauge fluctuations associated with fractionalization,  without the need to invoke a large-N expansion.  While our model calculation depends on a specific model, the results exhibit a new class of topological superconductivity that may occur under a wide range of circumstances. The model state we have uncovered involves $Z_2$ fractionalized order: the pairing between electrons and a $Z_2$ spin liquid with gapless spin excitations.  A fascinating aspect  of this topological superconductivity, is the spin-1/2 character of the order parameter, for unlike conventional Cooper pairing, the order parameter transforms under a $S=1/2$, double group representation and is consequently subject to Kramer's theorem, allowing for a spontaneous time-reversal breaking chiral superconductor, developing via a single phase transition,  independently of the underlying lattice. We remark that although our mean-field calculations suggest the existence of novel charge-$e$ topological superconductivity, fluctuations beyond mean-field theory may as well lead to other competing phases. Therefore, complementary numerical methods are necessary to justify the existence of the proposed phase, which is beyond the scope of this work and interesting for future studies.

We end by returning to our initial motivation, speculating on whether this novel type of order could develop in heavy fermion materials such as UTe$_2$.  Although our model calculation relies on a pre-existing spin liquid, which is not proved to exist in UTe$_2$ and hence our model is not directly applicable, it may also be possible to enter this novel superconducting state from a heavy Fermion liquid\cite{aaditya24}. 
As mentioned in the introduction, early measurements on superconducting UTe$_2$ suggested that the condensate  breaks time-reversal symmetry, producing a Kerr rotation, with STM evidence for chiral edge states.  More recent measurements, on improved samples have not reproduced the early Kerr rotation results, but chiral edge states are still believed to still be present. 
 A one-stage transition into a chiral topological superconductor is prohibited in orthorhombic Cooper-paired superconductors, as there are no two dimensional triplet representations. It would thus be interesting to revisit the question of time-reversal symmetry breaking in this material, for a single stage transition into a chiral state would be strong evidence for beyond BCS pairing.  
 
\section*{Acknowledgement}
This work was supported by 
the U.S. National Science Foundation grant DMR-1830707. Part of the work by Z.Z. was financially supported by the National Science Foundation, 
Quantum Leap Challenge Institute for Hybrid Quantum Architectures and Networks Grant No. OMA-2016136.
We gratefully acknowledge discussions
with Aaditya Panigrahi and Alexei Tsvelik.
\clearpage
\appendix
\begin{widetext}
\section{Transformation of order parameters under symmetry operations}\label{App:A}
In this section, we present how the order parameters transform under the time-reversal operation $\mathcal{\tilde T}$, the chiral operation $\mathcal{S}$ and the spin-lattice rotation $\mathcal{C}$ defined in the main text.

We first consider the effects of $\mathcal{S}$ and $\mathcal{\tilde{T}}$. As the original full Hamiltonian is invariant under the defined transformations $\mathcal{S}$ and $\mathcal{\tilde{T}}$, it suffices to investigate how the mean-field Hamiltonian transforms under these symmetry operations. As in the main text, the factorized mean-field Hamiltonian is given by
\begin{equation}
    H_{K1}=\sum_j\left\{[(c_j^\dagger \Vec{\sigma} V_j)\cdot\Vec{\chi}_j +\text{H.c.}]+2\frac{|V_j|^2}{J_1}\right\}, \label{HK1MFApp}
\end{equation}
\begin{equation}\label{HK2MFApp}
    H_{K2}=\sum_{\langle ij\rangle}\left\{ [(c_i^\dagger (\Vec{\sigma}\cdot \Vec{\chi}_j) V_{ji})+\text{H.c.}] + 2\frac{| V_{ij}|^2}{J_2} + (i\leftrightarrow j)\right\}. 
\end{equation}
It is straightforward to show that under the chiral operation $\mathcal{S}$
\begin{equation}
    \mathcal{S} (c_j^\dagger \Vec{\sigma} V_j)\cdot\Vec{\chi}_j\mathcal{S}^{-1}=(c_j^T \vec{\sigma}^* V_j^*)\cdot\vec{\chi}_j=(c_{j\alpha}\vec{\sigma}^*_{\alpha\beta}V_{j\beta}^*)\cdot \vec{\chi}_j=(V_{j\beta}^*\vec{\sigma}_{\beta\alpha}c_{j\alpha})\cdot\vec{\chi}_j=-\vec{\chi}_j\cdot(V_j^\dagger \vec{\sigma} c_j).
\end{equation}
Compare it with Eq. (\ref{HK1MFApp}) one concludes that under $\mathcal{S}$, $V_j\rightarrow -V_j$. Similarly one obtains that under $\mathcal{S}$, $V_{ij}\rightarrow V_{ij}$ where the extra minus sign is due to the $Z_2$ gauge transformation $\mathcal{G}$. Under the time-reversal operation $\mathcal{\tilde T}$, we have
\begin{equation}
    \mathcal{S} (c_j^\dagger \Vec{\sigma} V_j)\cdot\Vec{\chi}_j\mathcal{S}^{-1}=c_j^\dagger (-i\sigma_y \vec{\sigma^*}V_j^*)\cdot \vec{\chi_j}=(c_j^\dagger \vec{\sigma} i\sigma_y V_j^*)\cdot \vec{\chi}_j
\end{equation}
where we have used $i\sigma_y \vec{\sigma} i\sigma_y=\vec{\sigma}^T$. This indicates that under $\mathcal{\tilde T}$, $V_j \rightarrow i\sigma_y V_j^*$ and $V_{ij}=-i\sigma_y V_{ij}^*$.

The spin-lattice rotation $\mathcal{C}$ is a combination of the spin-rotation $\mathcal{U}$, the lattice rotation $\mathcal{R}$ and the $Z_2$ gauge transformation ${\mathcal G}$. Under the spin-rotation over angle $\pi/3$, the Majorana fermions in the spin liquid transform as a SO(3) vector
\begin{equation}
    \vec{\chi}\rightarrow R_z^{-1}(\frac \pi 3) \vec{\chi},
\end{equation}
while the conduction electrons transform as
\begin{equation}
    c \rightarrow e^{i\frac{\pi}{6}\sigma_z}c .
\end{equation}
Neglecting the multiplicative factor, one obtains that the on-site spinor order parameter $V_j$ transforms as
\begin{equation}
    V_j=(\Vec{\chi}_j\cdot \Vec{\sigma}) c_j\rightarrow (R_z^{-1}(\frac \pi 3) \Vec{\chi}_j\cdot \Vec{\sigma}) e^{i\frac{\pi}{6}\sigma_z}c_j=e^{i\frac{\pi}{6}\sigma_z}(\Vec{\chi}_j\cdot \Vec{\sigma}) c_j=e^{i\frac{\pi}{6}\sigma_z} V_j,
\end{equation}
where we have used the identity $e^{-i\frac{\pi}{6}\sigma_z}\vec{\sigma}e^{i\frac{\pi}{6}\sigma_z}=R_z(\pi/3)\vec{\sigma}$. Now we consider how the ansatz we adopt in the main text transforms under $\mathcal{C}$. As $v_j$ is uniform and involves fermions on the same site, $\mathcal{R}$ and ${\mathcal G}$ have no effects. Therefore 
\begin{equation}
    v_j=\left(\begin{array}{c}
         1 \\
          0
    \end{array}\right)\rightarrow e^{i\frac{\pi}{6}\sigma_z}\left(\begin{array}{c}
         1 \\
          0
    \end{array}\right)=e^{i\frac{\pi}{6}}v_j.
\end{equation}
For $v_{ij}$, the lattice rotation results in $\theta_{ij}\rightarrow \theta_{ij}+\pi/3$, while the gauge transformation ${\mathcal G}$ gives an extra minus sign, so
\begin{equation}
    v_{ij}=\left(\begin{array}{c}
         -\sin \frac{\gamma}{2}e^{-i\phi}e^{3i\theta_{ij}} \\
         \cos \frac{\gamma}{2}e^{4i\theta_{ij}}
    \end{array}\right)\rightarrow -e^{i\frac{\pi}{6}\sigma_z}\left(\begin{array}{c}
         \sin \frac{\gamma}{2}e^{-i\phi}e^{3i\theta_{ij}} \\
         -\cos \frac{\gamma}{2}e^{4i\theta_{ij}}e^{i\pi/3}
    \end{array}\right)=e^{i\frac{\pi}{6}}v_{ij}.
\end{equation}
We hence have proved that $v_j$ and $v_{ij}$ transform in the same 1-D irreducible representation.

\section{Full mean-field Hamiltonians}\label{App:B}
In this section, we present the full mean-field Hamiltonians in terms of Majorana fermions.

With the mean-field ansatz and Majorana representation in the main text, $H_C$, $H_R$ and $H_K$ can be rewritten as
\begin{equation}
    H_C=-it\sum_{\langle ij\rangle} c_{i}^\dagger c_{j }+\text{H.c.}=-it\sum_{\langle ij\rangle}\sum_{\alpha=0}^3 c_{i}^\alpha c_{j}^\alpha,  \label{FermiseaMajoranaH}
\end{equation}
\begin{align}\nonumber
    H_R&=-\lambda_R\sum_{\langle ij\rangle} c_{i}^\dagger \left[(\Vec{\sigma}\times \Vec{d}_{ij}) \cdot \hat{z}\right]c_{j }+\text{H.c.}\\
    &=-i\lambda_R\sum_{\langle ij\rangle}\left[d_{ij}^y\left(c_i^0c_j^1-c_i^1c_j^0+c_i^3c_j^2-c_i^2c_j^3\right)
    -d_{ij}^x\left(c_i^0c_j^2-c_i^2c_j^0+c_i^1c_j^3-c_i^3c_j^1\right)\right],
\end{align}
\begin{equation}
    H_{K1}=\frac{1}{2}\sum_{j} J_1\Vec{\sigma}_j\cdot(c_{j}^\dagger \Vec{\sigma} c_{j })=-iW_1\sum_j (\vec{c}_j\cdot \vec{\chi}_j)+\frac{2NW_1^2}{J_1},
\end{equation}

\begin{align} \label{NNKondoMajoranaH}
    H_{K2}=\frac{1}{2}\sum_{\langle ij\rangle} &J_{2}\Vec{\sigma}_i\cdot(c_{j}^\dagger \Vec{\sigma} c_{j })\\\nonumber
    =iW_2\sum_{\langle ij\rangle}&\left[\cos \frac{\gamma}{2}\sin(4\theta_{ij}+\beta)\left(c_j^0 \chi_i^1-c_j^3\chi_i^2+c_j^2\chi_i^3\right)-\cos \frac{\gamma}{2}\cos(4\theta_{ij}+\beta)\left(c_j^3 \chi_i^1+c_j^0\chi_i^2-c_j^1\chi_i^3\right)\right.\\ \nonumber
    &\left.+\sin \frac{\gamma}{2}\cos(3\theta_{ij}+\beta-\phi)\left(c_j^1 \chi_i^1+c_j^2\chi_i^2+c_j^3\chi_i^3\right)+\sin \frac{\gamma}{2}\sin(3\theta_{ij}+\beta-\phi)\left(c_j^2 \chi_i^1-c_j^1\chi_i^2-c_j^0\chi_i^3\right)\right]\\\nonumber
    &(i\leftrightarrow j)+\frac{6NW_2^2}{J_2},
\end{align}
where $N$ is the number of unit cells. Eqs.(\ref{FermiseaMajoranaH})-(\ref{NNKondoMajoranaH}) can be diagonalized by Fourier transforming the Majorana operators
\begin{equation}
    \chi_{\mathbf{k}\Lambda}^{\alpha\dagger}=\frac{1}{\sqrt{N}}\sum_j e^{i\mathbf{k}\cdot \mathbf{R}_j}\chi_{j\Lambda}^{\alpha},
\end{equation}
\begin{equation}
    c_{\mathbf{k}\Lambda}^{\alpha\dagger}=\frac{1}{\sqrt{N}}\sum_j e^{i\mathbf{k}\cdot \mathbf{R}_j}c_{j\Lambda}^\alpha,
\end{equation}
where $\mathbf{R}_j$ is the position of the unit cell and $\Lambda=A,B$ labels different sublattices. In momentum space, the Hamiltonian reads
\begin{equation}
    \tilde{H}_C=-it\sum_{\mathbf{k}\in \frac{1}{2}\text{BZ}}\sum_{\alpha=0}^3\left(A(\mathbf{k})c_{\mathbf{k}A}^{\alpha\dagger}c_{\mathbf{k}B}^{\alpha}-A^*(\mathbf{k})c_{\mathbf{k}B}^{\alpha\dagger}c_{\mathbf{k}A}^{\alpha}\right),
\end{equation}
\begin{equation}
    H_{YL}=iK\sum_{\mathbf{k}\in \frac{1}{2}\text{BZ}}\sum_{\alpha=1}^3\left(A(\mathbf{k})\chi_{\mathbf{k}A}^{\alpha\dagger}\chi_{\mathbf{k}B}^{\alpha}-A^*(\mathbf{k})\chi_{\mathbf{k}B}^{\alpha\dagger}\chi_{\mathbf{k}A}^{\alpha}\right),
\end{equation}
\begin{equation} \label{HRfull}
H_R=i\lambda_R\sum_{\mathbf{k}\in \frac{1}{2}\text{BZ}}
\left[B(\mathbf{k})\left(c_{\mathbf{k}A}^{0\dagger}c_{\mathbf{k}B}^{1}+c_{\mathbf{k}A}^{3\dagger}c_{\mathbf{k}B}^{2}-c_{\mathbf{k}A}^{1\dagger}c_{\mathbf{k}B}^{0}-c_{\mathbf{k}A}^{2\dagger}c_{\mathbf{k}B}^{3}\right)
+C(\mathbf{k})\left(c_{\mathbf{k}A}^{0\dagger}c_{\mathbf{k}B}^{2}+c_{\mathbf{k}A}^{1\dagger}c_{\mathbf{k}B}^{3}-c_{\mathbf{k}A}^{2\dagger}c_{\mathbf{k}B}^{0}-c_{\mathbf{k}A}^{3\dagger}c_{\mathbf{k}B}^{1}\right)\right]+\text{H.c.},
\end{equation}

\begin{align} \label{HKfull}
    H_{K}=-i\sum_{\mathbf{k}\in \frac{1}{2}\text{BZ}}&\left\{W_2\left[\tilde{B}(\mathbf{k})\left(c_{\mathbf{k}A}^{3\dagger}\chi_{\mathbf{k}B}^{1}+c_{\mathbf{k}A}^{0\dagger}\chi_{\mathbf{k}B}^{2}-c_{\mathbf{k}A}^{1\dagger}\chi_{\mathbf{k}B}^{3}\right)
    +\tilde{B}^*(\mathbf{k})\left(c_{\mathbf{k}B}^{3\dagger}\chi_{\mathbf{k}A}^{1}+c_{\mathbf{k}B}^{0\dagger}\chi_{\mathbf{k}A}^{2}-c_{\mathbf{k}B}^{1\dagger}\chi_{\mathbf{k}A}^{3}\right)\right]\right.\\\nonumber
    &-W_2\left[\tilde{C}(\mathbf{k})\left(c_{\mathbf{k}A}^{0\dagger}\chi_{\mathbf{k}B}^{1}-c_{\mathbf{k}A}^{3\dagger}\chi_{\mathbf{k}B}^{2}+c_{\mathbf{k}A}^{2\dagger}\chi_{\mathbf{k}B}^{3}\right)
    +\tilde{C}^*(\mathbf{k})\left(c_{\mathbf{k}B}^{0\dagger}\chi_{\mathbf{k}A}^{1}-c_{\mathbf{k}B}^{3\dagger}\chi_{\mathbf{k}A}^{2}+c_{\mathbf{k}B}^{2\dagger}\chi_{\mathbf{k}A}^{3}\right)\right]\\\nonumber
    &+W_2\sin \frac{\gamma}{2}\cos(\beta-\phi)\left(A(\mathbf{k})\sum_{\alpha=1}^3 c_{\mathbf{k}A}^{\alpha\dagger}\chi_{\mathbf{k}B}^{\alpha}-A^*(\mathbf{k})\sum_{\alpha=1}^3 c_{\mathbf{k}B}^{\alpha\dagger}\chi_{\mathbf{k}A}^{\alpha}\right)\\\nonumber
    &+W_2 \sin \frac{\gamma}{2}\sin(\beta-\phi)\left[A(\mathbf{k}) \left(c_{\mathbf{k}A}^{2\dagger}\chi_{\mathbf{k}B}^{1}-c_{\mathbf{k}A}^{1\dagger}\chi_{\mathbf{k}B}^{2}-c_{\mathbf{k}A}^{0\dagger}\chi_{\mathbf{k}B}^{3}\right)-A^*(\mathbf{k}) \left(c_{\mathbf{k}B}^{2\dagger}\chi_{\mathbf{k}A}^{1}-c_{\mathbf{k}B}^{1\dagger}\chi_{\mathbf{k}A}^{2}-c_{\mathbf{k}B}^{0\dagger}\chi_{\mathbf{k}A}^{3}\right)\right]\\\nonumber
    &\left.+W_1\sum_{\alpha=1}^3\left(c_{\mathbf{k}A}^{\alpha\dagger}\chi_{\mathbf{k}A}^{\alpha}+c_{\mathbf{k}B}^{\alpha\dagger}\chi_{\mathbf{k}B}^{\alpha}\right)+\text{H.c.}\right\},
\end{align}
where $A(\mathbf{k})=1+e^{-i\mathbf{k}\cdot \mathbf{a_1}}+e^{-i\mathbf{k}\cdot \mathbf{a_2}}$, $\tilde{B}(\mathbf{k})=\cos (\gamma/2)[\cos \beta+e^{-i\mathbf{k}\cdot \mathbf{a_1}}\cos (2\pi/3+\beta)+e^{-i\mathbf{k}\cdot \mathbf{a_2}}\cos (4\pi/3+\beta)]$, $\tilde{C}(\mathbf{k})=\cos (\gamma/2)[\sin \beta+e^{-i\mathbf{k}\cdot \mathbf{a_1}}\sin (2\pi/3+\beta)+e^{-i\mathbf{k}\cdot \mathbf{a_2}}\sin (4\pi/3+\beta)]$, $B(\mathbf{k})=\tilde{B}(\mathbf{k})\vert_{\gamma,\beta=0}$, $C(\mathbf{k})=\tilde{C}(\mathbf{k})\vert_{\gamma,\beta=0}$ and $\mathbf{a_1}=(\sqrt{3}/2,3/2)$, $\mathbf{a_2}=(-\sqrt{3}/2,3/2)$ are the Bravais lattice vectors. Here, we note that only half of the Brillouin zone (BZ) is summed over due to the redundancy $\chi_{\mathbf{k}\Lambda}^{\alpha\dagger}=\chi_{\mathbf{-k}\Lambda}^{\alpha}$ and $c_{\mathbf{k}\Lambda}^{\alpha\dagger}=c_{\mathbf{-k}\Lambda}^{\alpha}$. The self-consistency condition for the spinor order parameter $V_\Lambda$ at sublattice $\Lambda$ becomes
\begin{align}
    V_\Lambda=\frac{1}{N}\sum_{\mathbf{k}\in \frac{1}{2}\text{BZ}}\left(\begin{array}{c}
          V_\mathbf{k\Lambda\uparrow}\\
          V_\mathbf{k\Lambda\downarrow}
    \end{array}\right)=\frac{W_1}{\sqrt{2}}\left(\begin{array}{c}
         1 \\
          0
    \end{array}\right),
\end{align}
where 
\begin{equation}
    V_\mathbf{k\Lambda\uparrow}=-\frac{J_1}{2}\left[\sqrt{2}\left(X_{\mathbf{-k}\Lambda}c_{\mathbf{k}\Lambda,\downarrow}-c_{\mathbf{-k}\Lambda,\downarrow}X_{\mathbf{k}\Lambda}\right)+\chi_{\mathbf{k}\Lambda}^{3\dagger}c_{\mathbf{k}\Lambda,\uparrow}-c_{\mathbf{-k}\Lambda,\uparrow}\chi_{\mathbf{k}\Lambda}^{3}\right],
\end{equation}
\begin{equation}
    V_\mathbf{k\Lambda\downarrow}=-\frac{J_1}{2}\left[\sqrt{2}(X_{\mathbf{k}\Lambda}^\dagger c_{\mathbf{k}\Lambda,\uparrow}-c_{\mathbf{-k}\Lambda,\uparrow}X_{\mathbf{-k}\Lambda}^\dagger)-\chi_{\mathbf{k}\Lambda}^{3\dagger}c_{\mathbf{k}\Lambda,\downarrow}+c_{\mathbf{-k}\Lambda,\downarrow}\chi_{\mathbf{k}\Lambda}^{3}\right].
\end{equation}
Here we have defined $X_{\mathbf{k}\Lambda}=(\chi^1_{\mathbf{k}}-i\chi^2_{\mathbf{k}})/\sqrt{2}$ and used the relation $c_{\mathbf{k}\Lambda,\uparrow}=(c_{\mathbf{k}\Lambda}^0+ic_{\mathbf{k}\Lambda}^3)/\sqrt{2}$, $c_{\mathbf{k}\Lambda,\downarrow}=i(c_{\mathbf{k}\Lambda}^1+ic_{\mathbf{k}\Lambda}^2)/\sqrt{2}$. Similarly for the order parameter $V_{AB}\equiv V_{ij}$ where $ \mathbf{r}_i-\mathbf{r}_j=-\hat{y}$, we have
\begin{align}
    V_{AB}=\frac{1}{N}\sum_{\mathbf{k}\in \frac{1}{2}\text{BZ}}\left(\begin{array}{c}
          V_\mathbf{k,AB\uparrow}\\
          V_\mathbf{k,AB\downarrow}
    \end{array}\right)=\frac{W_2}{\sqrt{2}}\left(\begin{array}{c}
         0 \\
          1
    \end{array}\right),
\end{align}
where 
\begin{equation}
    V_\mathbf{k,AB\uparrow}=-\frac{J_2}{2}\left[\sqrt{2}\left(X_{\mathbf{-k}A}c_{\mathbf{k}B,\downarrow}-c_{\mathbf{-k}B,\downarrow}X_{\mathbf{k}A}\right)+\chi_{\mathbf{k}A}^{3\dagger}c_{\mathbf{k}B,\uparrow}-c_{\mathbf{-k}B,\uparrow}\chi_{\mathbf{k}A}^{3}\right],
\end{equation}
\begin{equation}
    V_\mathbf{k,AB\downarrow}=-\frac{J_2}{2}\left[\sqrt{2}(X_{\mathbf{k}A}^\dagger c_{\mathbf{k}B,\uparrow}-c_{\mathbf{-k}B,\uparrow}X_{\mathbf{-k}A}^\dagger)-\chi_{\mathbf{k}A}^{3\dagger}c_{\mathbf{k}B,\downarrow}+c_{\mathbf{-k}B,\downarrow}\chi_{\mathbf{k}A}^{3}\right].
\end{equation}
\section{Second-order perturbation calculations}\label{App:C}
In this section, we briefly apply the second-order perturbation theory to derive the low-energy Hamiltonian in the main text. For simplicity, we have assumed $t=K$.  
From Eq. (\ref{HRfull}) and Eq. (\ref{HKfull}), it is obvious that the first-order correction vanishes as there are no intraband matrix elements, so one has to calculate the second-order contribution of interband matrix elements. Near the high-symmetry momentum $K(K^\prime)=(\pm 4\pi/3\sqrt{3},0)$, the interband Hamiltonian is given by 
\begin{equation}
    \mathcal{P_L}H_{K2}(\mathbf{k})\mathcal{P_H}=-\frac{3iW_2}{2}\cos \frac{\gamma}{2}\times\left[c_{\mathbf{k}A}^{0\dagger}\left(\chi_{\mathbf{k}B}^2\pm i\chi_{\mathbf{k}B}^1\right)+c_{\mathbf{k}B}^{0\dagger}\left(\chi_{\mathbf{k}A}^2\mp i\chi_{\mathbf{k}A}^1\right)\right],
\end{equation}
\begin{equation}
    \mathcal{P_L}H_{R}(\mathbf{k})\mathcal{P_H}=\frac{3i\lambda_R}{2}\left[c_{\mathbf{k}A}^{0\dagger}\left(c_{\mathbf{k}B}^1\mp ic_{\mathbf{k}B}^2\right)+c_{\mathbf{k}B}^{0\dagger}\left(c_{\mathbf{k}A}^1\pm ic_{\mathbf{k}A}^2\right)\right],
\end{equation}
where $\mathcal{P_L}=|\mathcal{L}\rangle \langle \mathcal{L}|$ and $\mathcal{P_H}=|\mathcal{H}\rangle \langle \mathcal{H}|$ are the projection operators. From the second-order perturbation theory, the low-energy effective Hamiltonian is given by
\begin{equation}
    (H_\text{eff})_{mn}=\frac{1}{2}\sum_{l\in \mathcal{H}}\left(\frac{\langle m\vert H^\prime\vert l\rangle \langle l\vert H^\prime \vert n\rangle}{E_m-E_l}+\frac{\langle m\vert H^\prime\vert l\rangle \langle l\vert H^\prime \vert n\rangle}{E_n-E_l}\right),\label{HeffApp}
\end{equation}
where $H^\prime=H_R+H_{K2}$ is the perturbation. We shall restrict ourselves to the momentum at the high-symmetry point $K(K^\prime)=(\pm 4\pi/3\sqrt{3},0)$, at which the high-energy states are
\begin{equation}
    \vert l^\alpha_{\Lambda,\pm}\rangle=\frac{1}{\sqrt{2}}\left(c_{k,\Lambda}^{\alpha\dagger}\pm i\chi_{k,\Lambda}^{\alpha\dagger}\right)\vert 0\rangle
\end{equation}
with eigenenergies $E_l=\pm W_1$. One can show that Eq. (\ref{HeffApp}) vanishes to the order $O(W_2^2)$ or $O(\lambda_R^2)$, but is nonzero to the order $O(W_2 \lambda_R)$. The matrix elements at $k=(\pm 4\pi/3\sqrt{3},0)$ between $\vert k,A\rangle$ and $\vert k,B\rangle$ vanish while
\begin{align}
        \langle k,A\vert H_\text{eff}\vert k,A\rangle&=\sum_{l\in \mathcal{H}}\left(\frac{\langle k,A\vert H_{K2}\vert l\rangle \langle l\vert H_R \vert k,A\rangle}{-E_l}+\text{c.c.}\right)\\\nonumber
    &=\sum_{\alpha=1}^3\left( \frac{\langle k,A\vert H_{K2}\vert l_{B,+}^\alpha\rangle \langle l_{B,+}^\alpha\vert H_R \vert k,A\rangle}{-W_1}+\frac{\langle k,A\vert H_{K2}\vert l_{B,-}^\alpha\rangle \langle l_{B,-}^\alpha\vert H_R \vert k,A\rangle}{W_1}+\text{c.c.}\right)\\\nonumber
    &=\mp\frac{9W_2\lambda_R}{W_1}\cos \frac{\gamma}{2}\cos \beta,
\end{align}
\begin{align}
        \langle k,B\vert H_\text{eff}\vert k,B\rangle&=\sum_{l\in \mathcal{H}}\left(\frac{\langle k,B\vert H_{K2}\vert l\rangle \langle l\vert H_R \vert k,B\rangle}{-E_l}+\text{c.c.}\right)\\\nonumber
    &=\sum_{\alpha=1}^3\left( \frac{\langle k,B\vert H_{K2}\vert l_{A,+}^\alpha\rangle \langle l_{A,+}^\alpha\vert H_R \vert k,B\rangle}{-W_1}+\frac{\langle k,B\vert H_{K2}\vert l_{A,-}^\alpha\rangle \langle l_{A,-}^\alpha\vert H_R \vert k,B\rangle}{W_1}+\text{c.c.}\right)\\\nonumber
    &=\pm\frac{9W_2\lambda_R}{W_1}\cos \frac{\gamma}{2}\cos \beta,
\end{align}
where we have used Eq. (\ref{HRfull}) and Eq. (\ref{HKfull}).

\section{Mean-field diagram}\label{App:D}
\begin{figure*}[!ht]
    \subfigure[]{\label{fig:phasediagram1App}
    \includegraphics[width=0.32 \textwidth]{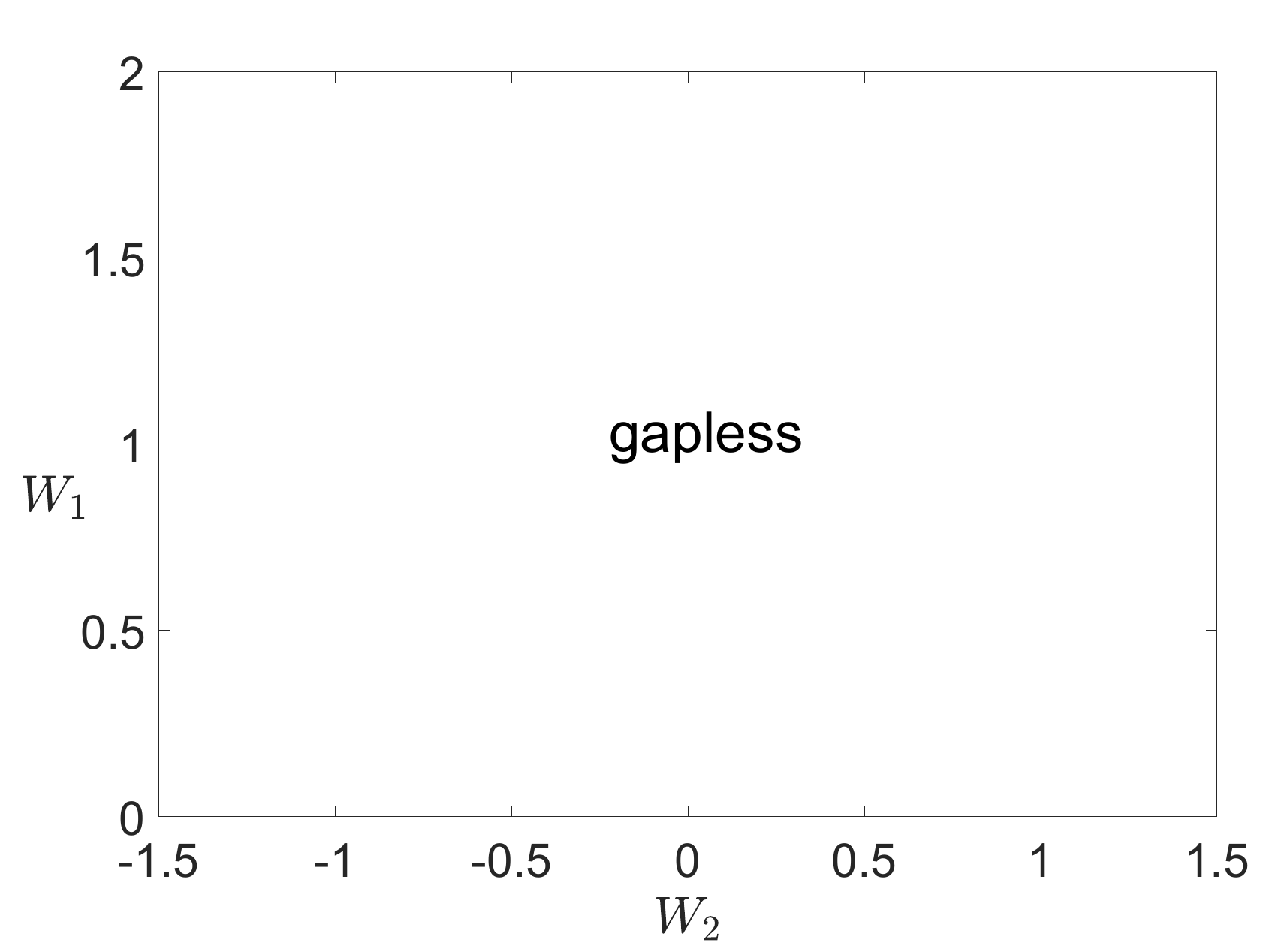}}
    \subfigure[]{\label{fig:phasediagram2App}
    \includegraphics[width=0.32 \textwidth]{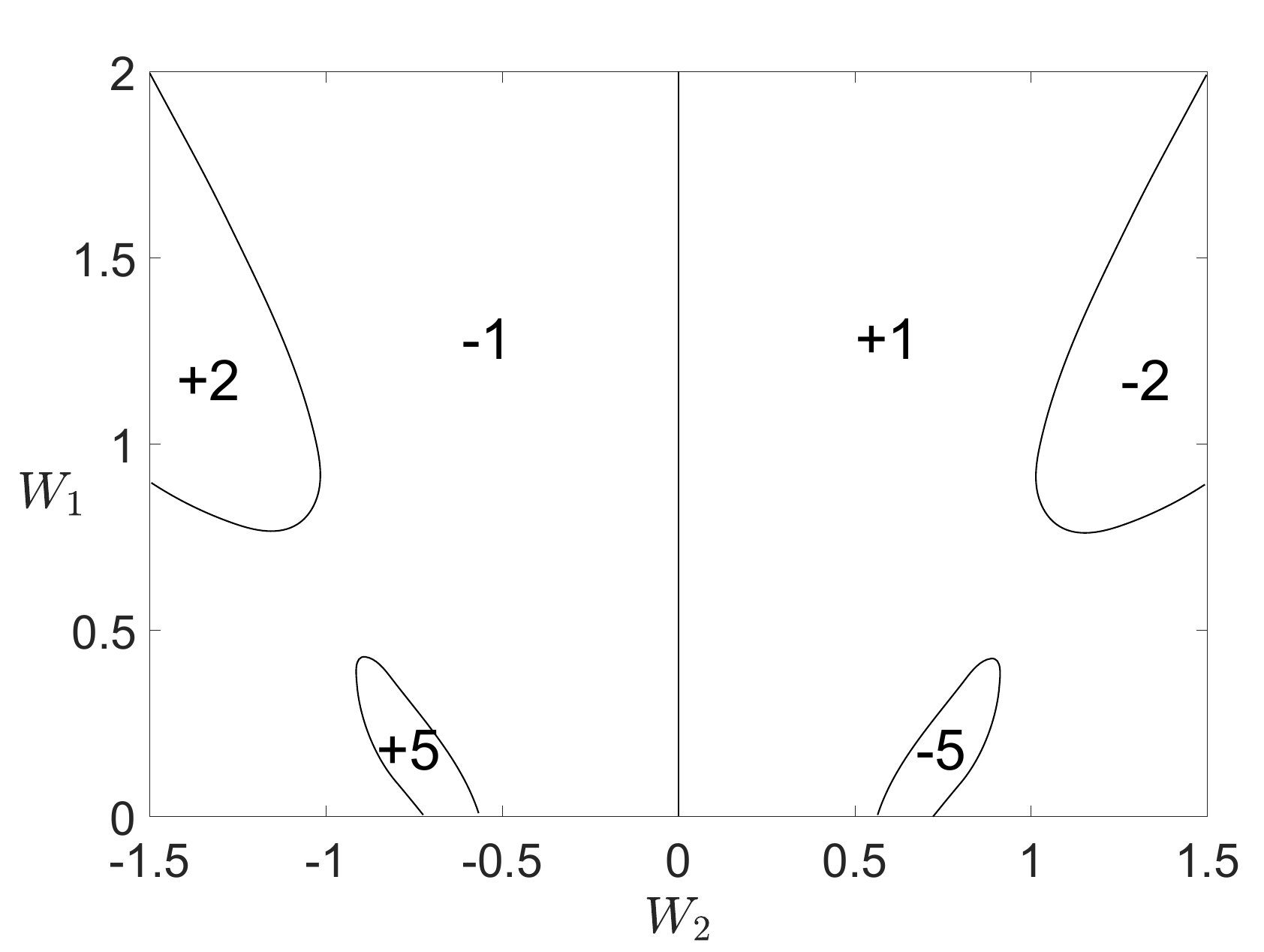}}  
        \subfigure[]{\label{fig:phasediagram3App}
    \includegraphics[width=0.32 \textwidth]{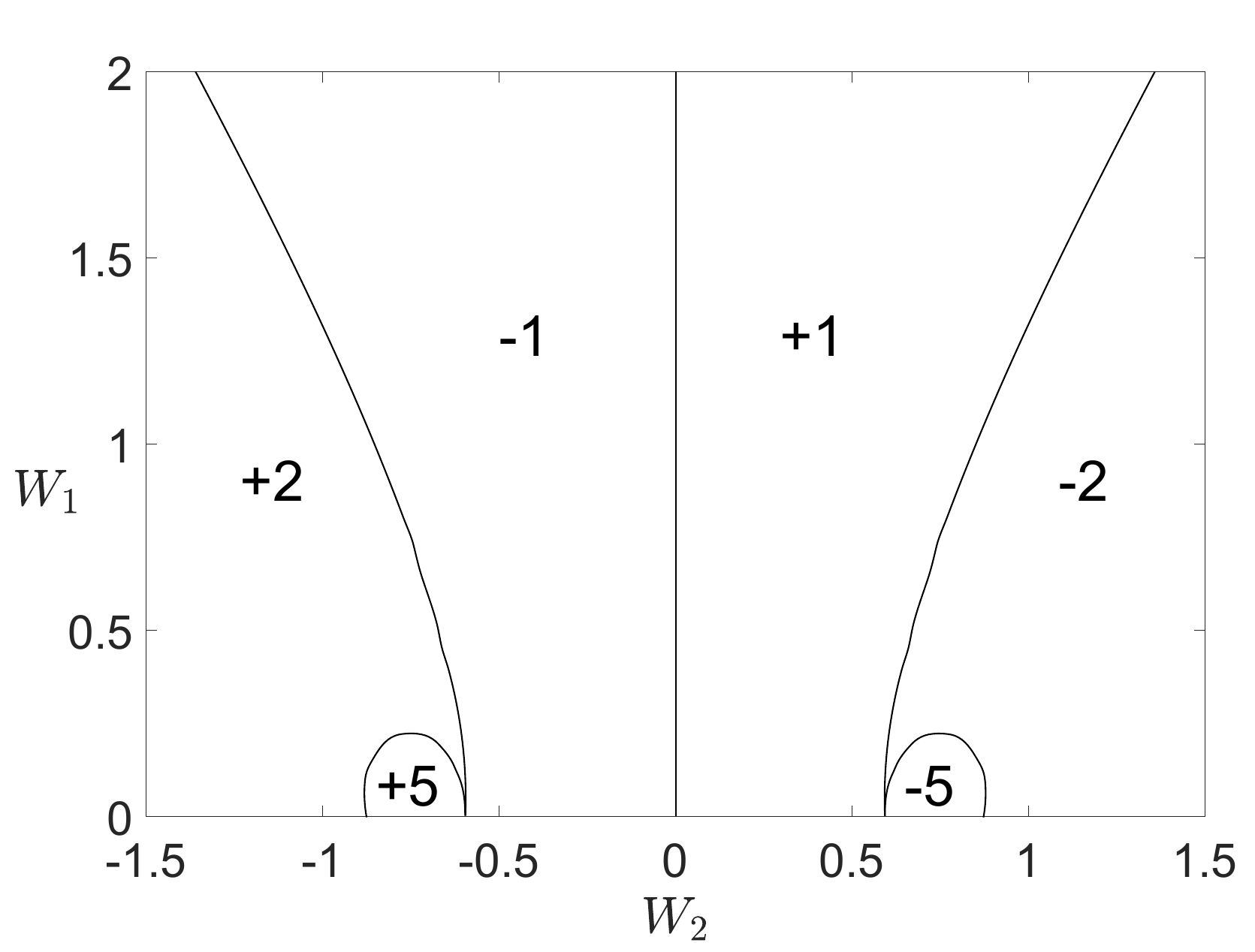}}  
        \subfigure[]{\label{fig:phasediagram4App}
    \includegraphics[width=0.32 \textwidth]{mu0lambda0_3_4.pdf}}
    \subfigure[]{\label{fig:phasediagram5App}
    \includegraphics[width=0.32 \textwidth]{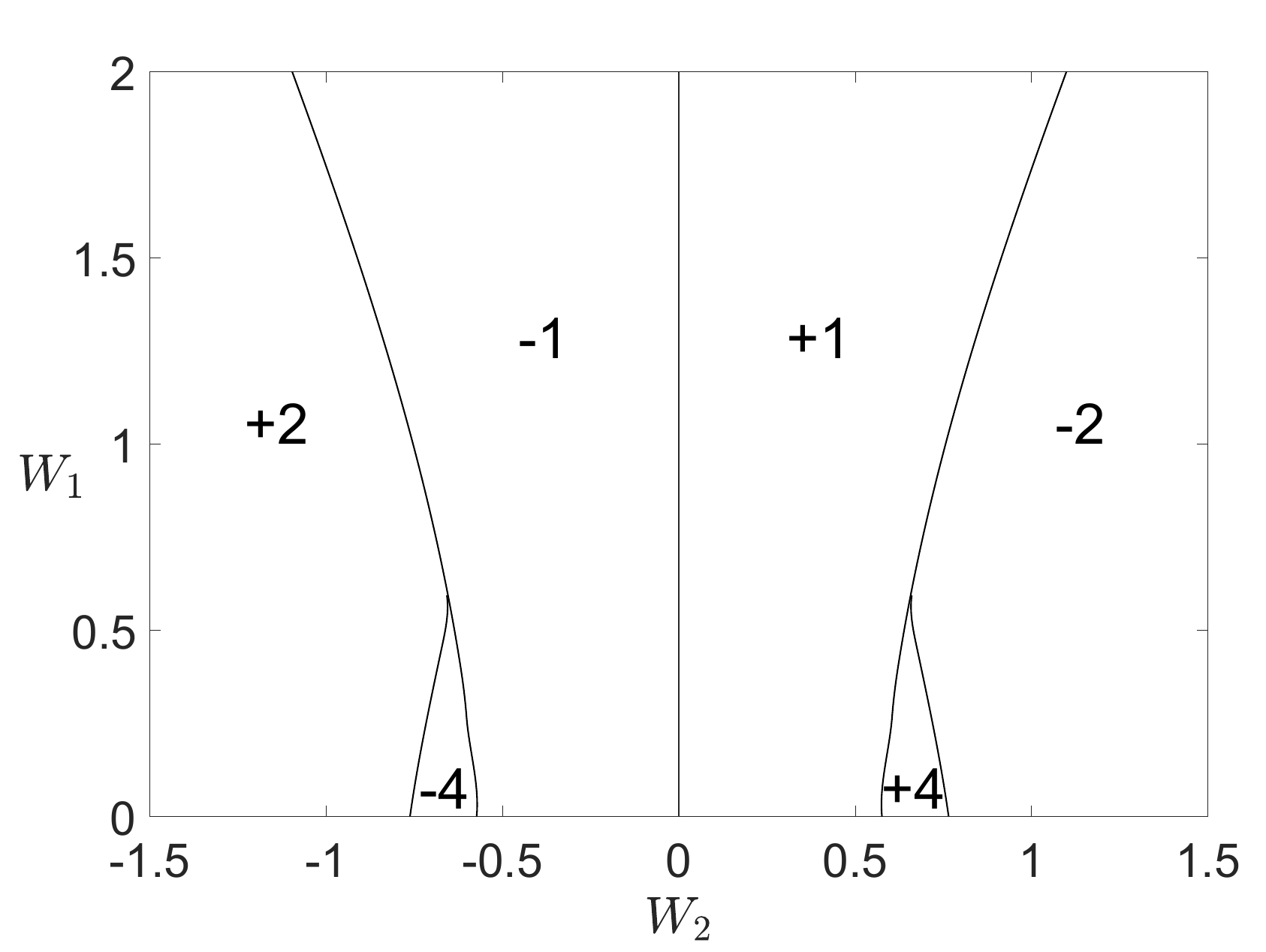}}  
        \subfigure[]{\label{fig:phasediagram6App}
    \includegraphics[width=0.32 \textwidth]{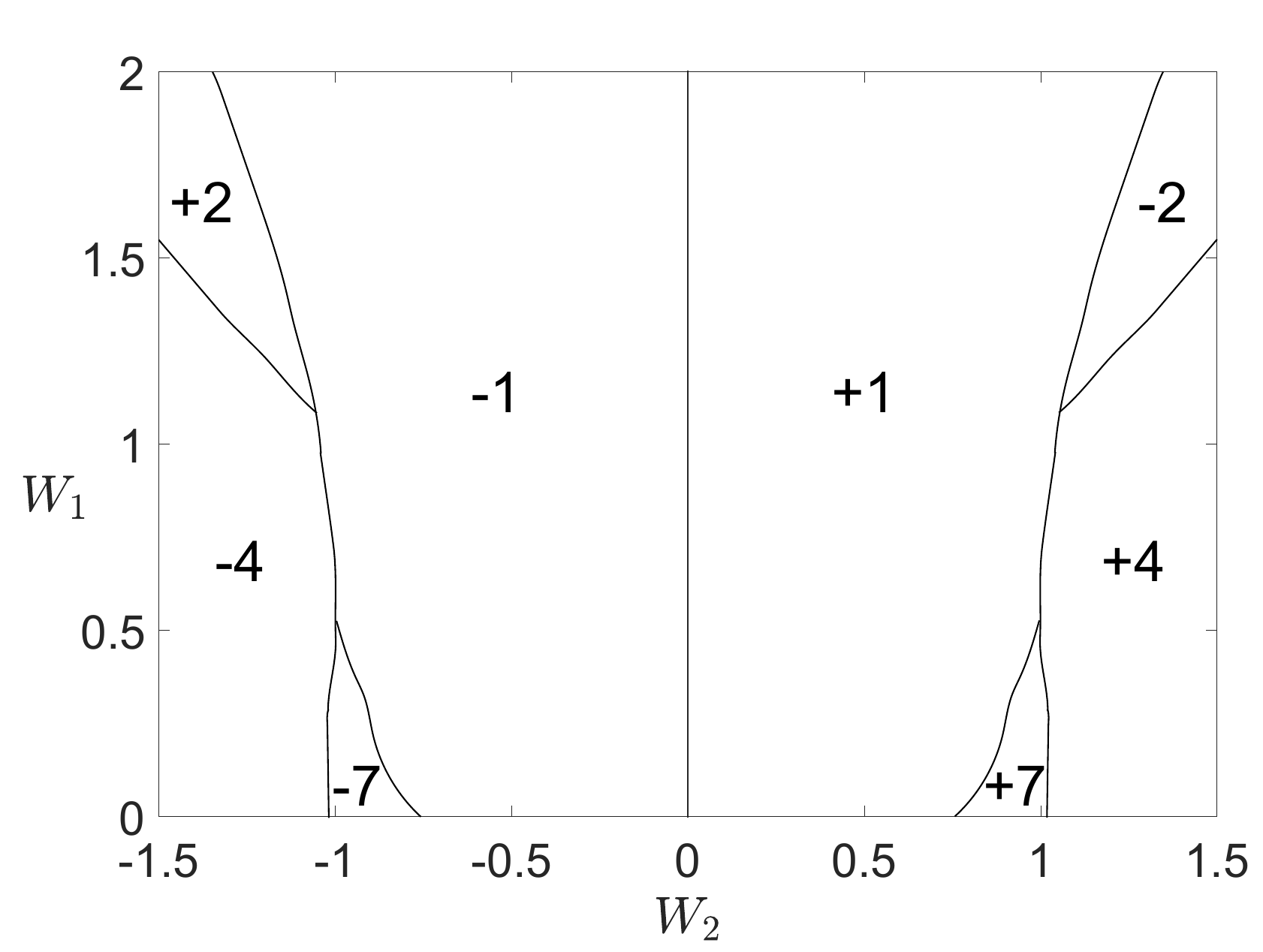}}  
    \caption{The mean-field phase diagram as a function of $W_1$ and $W_2$ when (a) $\gamma=-\pi$, (b) $\gamma=-2\pi/3$, (c) $\gamma=-\pi/3$, (d) $\gamma=0$, (e) $\gamma=\pi/3$, (f) $\gamma=2\pi/3$. The parameters used are: $\lambda_R=0.3, \mu=0$ and $t=K=1$. The number indicates the Chern number of the corresponding topological phase.}
   \label{fig:phasediagram0_0_3} 
\end{figure*}
Figure \ref{fig:phasediagram0_0_3} shows the mean-field phase diagram as a function of $W_1$ and $W_2$ for different $\gamma$s.

\end{widetext}

\clearpage
\bibliography{combined6}

\begin{thebibliography}{30}%
\makeatletter
\providecommand \@ifxundefined [1]{%
 \@ifx{#1\undefined}
}%
\providecommand \@ifnum [1]{%
 \ifnum #1\expandafter \@firstoftwo
 \else \expandafter \@secondoftwo
 \fi
}%
\providecommand \@ifx [1]{%
 \ifx #1\expandafter \@firstoftwo
 \else \expandafter \@secondoftwo
 \fi
}%
\providecommand \natexlab [1]{#1}%
\providecommand \enquote  [1]{``#1''}%
\providecommand \bibnamefont  [1]{#1}%
\providecommand \bibfnamefont [1]{#1}%
\providecommand \citenamefont [1]{#1}%
\providecommand \href@noop [0]{\@secondoftwo}%
\providecommand \href [0]{\begingroup \@sanitize@url \@href}%
\providecommand \@href[1]{\@@startlink{#1}\@@href}%
\providecommand \@@href[1]{\endgroup#1\@@endlink}%
\providecommand \@sanitize@url [0]{\catcode `\\12\catcode `\$12\catcode
  `\&12\catcode `\#12\catcode `\^12\catcode `\_12\catcode `\%12\relax}%
\providecommand \@@startlink[1]{}%
\providecommand \@@endlink[0]{}%
\providecommand \url  [0]{\begingroup\@sanitize@url \@url }%
\providecommand \@url [1]{\endgroup\@href {#1}{\urlprefix }}%
\providecommand \urlprefix  [0]{URL }%
\providecommand \Eprint [0]{\href }%
\providecommand \doibase [0]{https://doi.org/}%
\providecommand \selectlanguage [0]{\@gobble}%
\providecommand \bibinfo  [0]{\@secondoftwo}%
\providecommand \bibfield  [0]{\@secondoftwo}%
\providecommand \translation [1]{[#1]}%
\providecommand \BibitemOpen [0]{}%
\providecommand \bibitemStop [0]{}%
\providecommand \bibitemNoStop [0]{.\EOS\space}%
\providecommand \EOS [0]{\spacefactor3000\relax}%
\providecommand \BibitemShut  [1]{\csname bibitem#1\endcsname}%
\let\auto@bib@innerbib\@empty
\bibitem [{\citenamefont {Menth}\ \emph {et~al.}(1969)\citenamefont {Menth},
  \citenamefont {Buehler},\ and\ \citenamefont {Geballe}}]{menth_prl1968}%
  \BibitemOpen
  \bibfield  {author} {\bibinfo {author} {\bibfnamefont {A.}~\bibnamefont
  {Menth}}, \bibinfo {author} {\bibfnamefont {E.}~\bibnamefont {Buehler}},\
  and\ \bibinfo {author} {\bibfnamefont {T.}~\bibnamefont {Geballe}},\
  }\bibfield  {title} {\bibinfo {title} {{Magnetic and Semiconducting
  Properties of SmB$_{6}$}},\ }\href
  {https://doi.org/10.1103/physrevlett.22.295} {\bibfield  {journal} {\bibinfo
  {journal} {Physical review letters}\ }\textbf {\bibinfo {volume} {22}},\
  \bibinfo {pages} {295–297} (\bibinfo {year} {1969})}\BibitemShut {NoStop}%
\bibitem [{\citenamefont {Andres}\ \emph {et~al.}(1975)\citenamefont {Andres},
  \citenamefont {Graebner},\ and\ \citenamefont {Ott}}]{ott76}%
  \BibitemOpen
  \bibfield  {author} {\bibinfo {author} {\bibfnamefont {K.}~\bibnamefont
  {Andres}}, \bibinfo {author} {\bibfnamefont {J.}~\bibnamefont {Graebner}},\
  and\ \bibinfo {author} {\bibfnamefont {H.~R.}\ \bibnamefont {Ott}},\
  }\bibfield  {title} {\bibinfo {title} {{ { 4f-Virtual-Bound-State Formation
  in $CeAl_3$ at Low Temperatures}}},\ }\href
  {https://doi.org/10.1103/physrevlett.35.1779} {\bibfield  {journal} {\bibinfo
   {journal} {{P}hys. {R}ev. {L}ett.}\ }\textbf {\bibinfo {volume} {35}},\
  \bibinfo {pages} {1779} (\bibinfo {year} {1975})}\BibitemShut {NoStop}%
\bibitem [{\citenamefont {Steglich}\ \emph {et~al.}(1976)\citenamefont
  {Steglich}, \citenamefont {Aarts}, \citenamefont {Bredl}, \citenamefont
  {Leike}, \citenamefont {Franz},\ and\ \citenamefont {Sch{\"
  a}fer}}]{steglich}%
  \BibitemOpen
  \bibfield  {author} {\bibinfo {author} {\bibfnamefont {F.}~\bibnamefont
  {Steglich}}, \bibinfo {author} {\bibfnamefont {J.}~\bibnamefont {Aarts}},
  \bibinfo {author} {\bibfnamefont {C.~D.}\ \bibnamefont {Bredl}}, \bibinfo
  {author} {\bibfnamefont {W.}~\bibnamefont {Leike}}, \bibinfo {author}
  {\bibfnamefont {D.~E. M.~W.}\ \bibnamefont {Franz}},\ and\ \bibinfo {author}
  {\bibfnamefont {H.}~\bibnamefont {Sch{\" a}fer}},\ }\bibfield  {title}
  {\bibinfo {title} {{ { Superconductivity in the Presence of Strong Pauli
  Paramagnetism: $CeCu_ {2} Si_ {2}$}}},\ }\href
  {https://doi.org/10.1103/physrevlett.43.1892} {\bibfield  {journal} {\bibinfo
   {journal} {{P}hys. {R}ev. {L}ett}\ }\textbf {\bibinfo {volume} {43}},\
  \bibinfo {pages} {1892} (\bibinfo {year} {1976})}\BibitemShut {NoStop}%
\bibitem [{\citenamefont {Ott}\ \emph {et~al.}(1983)\citenamefont {Ott},
  \citenamefont {Rudigier}, \citenamefont {Fisk},\ and\ \citenamefont
  {Smith}}]{ube13a}%
  \BibitemOpen
  \bibfield  {author} {\bibinfo {author} {\bibfnamefont {H.~R.}\ \bibnamefont
  {Ott}}, \bibinfo {author} {\bibfnamefont {H.}~\bibnamefont {Rudigier}},
  \bibinfo {author} {\bibfnamefont {Z.}~\bibnamefont {Fisk}},\ and\ \bibinfo
  {author} {\bibfnamefont {J.~L.}\ \bibnamefont {Smith}},\ }\bibfield  {title}
  {\bibinfo {title} {{ { $UBe_{13}$: an unconventional actinide
  superconductor}}},\ }\href {https://doi.org/10.1103/physrevlett.50.1595}
  {\bibfield  {journal} {\bibinfo  {journal} {{P}hys. {R}ev. {L}ett}\ }\textbf
  {\bibinfo {volume} {50}},\ \bibinfo {pages} {1595} (\bibinfo {year}
  {1983})}\BibitemShut {NoStop}%
\bibitem [{\citenamefont {Kane}\ and\ \citenamefont
  {Mele}(2005{\natexlab{a}})}]{Kane2005_1}%
  \BibitemOpen
  \bibfield  {author} {\bibinfo {author} {\bibfnamefont {C.~L.}\ \bibnamefont
  {Kane}}\ and\ \bibinfo {author} {\bibfnamefont {E.~J.}\ \bibnamefont
  {Mele}},\ }\bibfield  {title} {\bibinfo {title} {{Quantum Spin Hall Effect in
  Graphene}},\ }\href {https://doi.org/10.1103/PhysRevLett.95.226801}
  {\bibfield  {journal} {\bibinfo  {journal} {Phys. Rev. Lett.}\ }\textbf
  {\bibinfo {volume} {95}},\ \bibinfo {pages} {226801} (\bibinfo {year}
  {2005}{\natexlab{a}})}\BibitemShut {NoStop}%
\bibitem [{\citenamefont {Kane}\ and\ \citenamefont
  {Mele}(2005{\natexlab{b}})}]{Kane2005_2}%
  \BibitemOpen
  \bibfield  {author} {\bibinfo {author} {\bibfnamefont {C.~L.}\ \bibnamefont
  {Kane}}\ and\ \bibinfo {author} {\bibfnamefont {E.~J.}\ \bibnamefont
  {Mele}},\ }\bibfield  {title} {\bibinfo {title} {{${Z}_{2}$ Topological Order
  and the Quantum Spin Hall Effect}},\ }\href
  {https://doi.org/10.1103/PhysRevLett.95.146802} {\bibfield  {journal}
  {\bibinfo  {journal} {Phys. Rev. Lett.}\ }\textbf {\bibinfo {volume} {95}},\
  \bibinfo {pages} {146802} (\bibinfo {year} {2005}{\natexlab{b}})}\BibitemShut
  {NoStop}%
\bibitem [{\citenamefont {Dzero}\ \emph {et~al.}(2010)\citenamefont {Dzero},
  \citenamefont {Sun}, \citenamefont {Galitski},\ and\ \citenamefont
  {Coleman}}]{Dzero2010}%
  \BibitemOpen
  \bibfield  {author} {\bibinfo {author} {\bibfnamefont {M.}~\bibnamefont
  {Dzero}}, \bibinfo {author} {\bibfnamefont {K.}~\bibnamefont {Sun}}, \bibinfo
  {author} {\bibfnamefont {V.}~\bibnamefont {Galitski}},\ and\ \bibinfo
  {author} {\bibfnamefont {P.}~\bibnamefont {Coleman}},\ }\bibfield  {title}
  {\bibinfo {title} {{Topological Kondo Insulators}},\ }\href
  {https://doi.org/10.1103/PhysRevLett.104.106408} {\bibfield  {journal}
  {\bibinfo  {journal} {Phys. Rev. Lett.}\ }\textbf {\bibinfo {volume} {104}},\
  \bibinfo {pages} {106408} (\bibinfo {year} {2010})}\BibitemShut {NoStop}%
\bibitem [{\citenamefont {Alexandrov}\ \emph {et~al.}(2013)\citenamefont
  {Alexandrov}, \citenamefont {Dzero},\ and\ \citenamefont
  {Coleman}}]{Alexandrov2013}%
  \BibitemOpen
  \bibfield  {author} {\bibinfo {author} {\bibfnamefont {V.}~\bibnamefont
  {Alexandrov}}, \bibinfo {author} {\bibfnamefont {M.}~\bibnamefont {Dzero}},\
  and\ \bibinfo {author} {\bibfnamefont {P.}~\bibnamefont {Coleman}},\
  }\bibfield  {title} {\bibinfo {title} {{Cubic Topological Kondo
  Insulators}},\ }\href {https://doi.org/10.1103/PhysRevLett.111.226403}
  {\bibfield  {journal} {\bibinfo  {journal} {Phys. Rev. Lett.}\ }\textbf
  {\bibinfo {volume} {111}},\ \bibinfo {pages} {226403} (\bibinfo {year}
  {2013})}\BibitemShut {NoStop}%
\bibitem [{\citenamefont {Dzero}\ \emph {et~al.}(2016)\citenamefont {Dzero},
  \citenamefont {Xia}, \citenamefont {Galitski},\ and\ \citenamefont
  {Coleman}}]{Dzero2016}%
  \BibitemOpen
  \bibfield  {author} {\bibinfo {author} {\bibfnamefont {M.}~\bibnamefont
  {Dzero}}, \bibinfo {author} {\bibfnamefont {J.}~\bibnamefont {Xia}}, \bibinfo
  {author} {\bibfnamefont {V.}~\bibnamefont {Galitski}},\ and\ \bibinfo
  {author} {\bibfnamefont {P.}~\bibnamefont {Coleman}},\ }\bibfield  {title}
  {\bibinfo {title} {{Topological kondo insulators}},\ }\href
  {https://doi.org/10.1146/annurev-conmatphys-031214-014749} {\bibfield
  {journal} {\bibinfo  {journal} {Annual Review of Condensed Matter Physics}\
  }\textbf {\bibinfo {volume} {7}},\ \bibinfo {pages} {249} (\bibinfo {year}
  {2016})}\BibitemShut {NoStop}%
\bibitem [{\citenamefont {Ran}\ \emph {et~al.}(2019{\natexlab{a}})\citenamefont
  {Ran}, \citenamefont {Eckberg}, \citenamefont {Ding}, \citenamefont
  {Furukawa}, \citenamefont {Metz}, \citenamefont {Saha}, \citenamefont {Liu},
  \citenamefont {Zic}, \citenamefont {Kim}, \citenamefont {Paglione} \emph
  {et~al.}}]{Ran2019}%
  \BibitemOpen
  \bibfield  {author} {\bibinfo {author} {\bibfnamefont {S.}~\bibnamefont
  {Ran}}, \bibinfo {author} {\bibfnamefont {C.}~\bibnamefont {Eckberg}},
  \bibinfo {author} {\bibfnamefont {Q.-P.}\ \bibnamefont {Ding}}, \bibinfo
  {author} {\bibfnamefont {Y.}~\bibnamefont {Furukawa}}, \bibinfo {author}
  {\bibfnamefont {T.}~\bibnamefont {Metz}}, \bibinfo {author} {\bibfnamefont
  {S.~R.}\ \bibnamefont {Saha}}, \bibinfo {author} {\bibfnamefont {I.-L.}\
  \bibnamefont {Liu}}, \bibinfo {author} {\bibfnamefont {M.}~\bibnamefont
  {Zic}}, \bibinfo {author} {\bibfnamefont {H.}~\bibnamefont {Kim}}, \bibinfo
  {author} {\bibfnamefont {J.}~\bibnamefont {Paglione}}, \emph {et~al.},\
  }\bibfield  {title} {\bibinfo {title} {Nearly ferromagnetic spin-triplet
  superconductivity},\ }\href {https://doi.org/10.1126/science.aav8645}
  {\bibfield  {journal} {\bibinfo  {journal} {Science}\ }\textbf {\bibinfo
  {volume} {365}},\ \bibinfo {pages} {684} (\bibinfo {year}
  {2019}{\natexlab{a}})}\BibitemShut {NoStop}%
\bibitem [{\citenamefont {Ran}\ \emph {et~al.}(2019{\natexlab{b}})\citenamefont
  {Ran}, \citenamefont {Liu}, \citenamefont {Eo}, \citenamefont {Campbell},
  \citenamefont {Neves}, \citenamefont {Fuhrman}, \citenamefont {Saha},
  \citenamefont {Eckberg}, \citenamefont {Kim}, \citenamefont {Graf} \emph
  {et~al.}}]{Ran2019_2}%
  \BibitemOpen
  \bibfield  {author} {\bibinfo {author} {\bibfnamefont {S.}~\bibnamefont
  {Ran}}, \bibinfo {author} {\bibfnamefont {I.-L.}\ \bibnamefont {Liu}},
  \bibinfo {author} {\bibfnamefont {Y.~S.}\ \bibnamefont {Eo}}, \bibinfo
  {author} {\bibfnamefont {D.~J.}\ \bibnamefont {Campbell}}, \bibinfo {author}
  {\bibfnamefont {P.~M.}\ \bibnamefont {Neves}}, \bibinfo {author}
  {\bibfnamefont {W.~T.}\ \bibnamefont {Fuhrman}}, \bibinfo {author}
  {\bibfnamefont {S.~R.}\ \bibnamefont {Saha}}, \bibinfo {author}
  {\bibfnamefont {C.}~\bibnamefont {Eckberg}}, \bibinfo {author} {\bibfnamefont
  {H.}~\bibnamefont {Kim}}, \bibinfo {author} {\bibfnamefont {D.}~\bibnamefont
  {Graf}}, \emph {et~al.},\ }\bibfield  {title} {\bibinfo {title} {{Extreme
  magnetic field-boosted superconductivity}},\ }\href
  {https://doi.org/10.1038/s41567-019-0670-x} {\bibfield  {journal} {\bibinfo
  {journal} {Nature Physics}\ }\textbf {\bibinfo {volume} {15}},\ \bibinfo
  {pages} {1250} (\bibinfo {year} {2019}{\natexlab{b}})}\BibitemShut {NoStop}%
\bibitem [{\citenamefont {Aoki}\ \emph {et~al.}(2019)\citenamefont {Aoki},
  \citenamefont {Nakamura}, \citenamefont {Honda}, \citenamefont {Li},
  \citenamefont {Homma}, \citenamefont {Shimizu}, \citenamefont {Sato},
  \citenamefont {Knebel}, \citenamefont {Brison}, \citenamefont {Pourret} \emph
  {et~al.}}]{Aoki2019}%
  \BibitemOpen
  \bibfield  {author} {\bibinfo {author} {\bibfnamefont {D.}~\bibnamefont
  {Aoki}}, \bibinfo {author} {\bibfnamefont {A.}~\bibnamefont {Nakamura}},
  \bibinfo {author} {\bibfnamefont {F.}~\bibnamefont {Honda}}, \bibinfo
  {author} {\bibfnamefont {D.}~\bibnamefont {Li}}, \bibinfo {author}
  {\bibfnamefont {Y.}~\bibnamefont {Homma}}, \bibinfo {author} {\bibfnamefont
  {Y.}~\bibnamefont {Shimizu}}, \bibinfo {author} {\bibfnamefont {Y.~J.}\
  \bibnamefont {Sato}}, \bibinfo {author} {\bibfnamefont {G.}~\bibnamefont
  {Knebel}}, \bibinfo {author} {\bibfnamefont {J.-P.}\ \bibnamefont {Brison}},
  \bibinfo {author} {\bibfnamefont {A.}~\bibnamefont {Pourret}}, \emph
  {et~al.},\ }\bibfield  {title} {\bibinfo {title} {{Unconventional
  superconductivity in heavy fermion UTe$_2$}},\ }\href
  {https://doi.org/10.7566/JPSJ.88.043702} {\bibfield  {journal} {\bibinfo
  {journal} {Journal of the Physical society of Japan}\ }\textbf {\bibinfo
  {volume} {88}},\ \bibinfo {pages} {043702} (\bibinfo {year}
  {2019})}\BibitemShut {NoStop}%
\bibitem [{\citenamefont {Aoki}\ \emph {et~al.}(2022)\citenamefont {Aoki},
  \citenamefont {Brison}, \citenamefont {Flouquet}, \citenamefont {Ishida},
  \citenamefont {Knebel}, \citenamefont {Tokunaga},\ and\ \citenamefont
  {Yanase}}]{Aoki2022}%
  \BibitemOpen
  \bibfield  {author} {\bibinfo {author} {\bibfnamefont {D.}~\bibnamefont
  {Aoki}}, \bibinfo {author} {\bibfnamefont {J.-P.}\ \bibnamefont {Brison}},
  \bibinfo {author} {\bibfnamefont {J.}~\bibnamefont {Flouquet}}, \bibinfo
  {author} {\bibfnamefont {K.}~\bibnamefont {Ishida}}, \bibinfo {author}
  {\bibfnamefont {G.}~\bibnamefont {Knebel}}, \bibinfo {author} {\bibfnamefont
  {Y.}~\bibnamefont {Tokunaga}},\ and\ \bibinfo {author} {\bibfnamefont
  {Y.}~\bibnamefont {Yanase}},\ }\bibfield  {title} {\bibinfo {title}
  {{Unconventional superconductivity in UTe$_2$}},\ }\href
  {https://doi.org/10.1088/1361-648X/ac5863} {\bibfield  {journal} {\bibinfo
  {journal} {Journal of Physics: Condensed Matter}\ }\textbf {\bibinfo {volume}
  {34}},\ \bibinfo {pages} {243002} (\bibinfo {year} {2022})}\BibitemShut
  {NoStop}%
\bibitem [{\citenamefont {Hayes}\ \emph {et~al.}(2021)\citenamefont {Hayes},
  \citenamefont {Wei}, \citenamefont {Metz}, \citenamefont {Zhang},
  \citenamefont {Eo}, \citenamefont {Ran}, \citenamefont {Saha}, \citenamefont
  {Collini}, \citenamefont {Butch}, \citenamefont {Agterberg} \emph
  {et~al.}}]{Hayes2021}%
  \BibitemOpen
  \bibfield  {author} {\bibinfo {author} {\bibfnamefont {I.}~\bibnamefont
  {Hayes}}, \bibinfo {author} {\bibfnamefont {D.~S.}\ \bibnamefont {Wei}},
  \bibinfo {author} {\bibfnamefont {T.}~\bibnamefont {Metz}}, \bibinfo {author}
  {\bibfnamefont {J.}~\bibnamefont {Zhang}}, \bibinfo {author} {\bibfnamefont
  {Y.~S.}\ \bibnamefont {Eo}}, \bibinfo {author} {\bibfnamefont
  {S.}~\bibnamefont {Ran}}, \bibinfo {author} {\bibfnamefont {S.}~\bibnamefont
  {Saha}}, \bibinfo {author} {\bibfnamefont {J.}~\bibnamefont {Collini}},
  \bibinfo {author} {\bibfnamefont {N.}~\bibnamefont {Butch}}, \bibinfo
  {author} {\bibfnamefont {D.}~\bibnamefont {Agterberg}}, \emph {et~al.},\
  }\bibfield  {title} {\bibinfo {title} {{Multicomponent superconducting order
  parameter in UTe$_2$}},\ }\href {https://doi.org/10.1126/science.abb0272}
  {\bibfield  {journal} {\bibinfo  {journal} {Science}\ }\textbf {\bibinfo
  {volume} {373}},\ \bibinfo {pages} {797} (\bibinfo {year}
  {2021})}\BibitemShut {NoStop}%
\bibitem [{\citenamefont {Bae}\ \emph {et~al.}(2021)\citenamefont {Bae},
  \citenamefont {Kim}, \citenamefont {Eo}, \citenamefont {Ran}, \citenamefont
  {Liu}, \citenamefont {Fuhrman}, \citenamefont {Paglione}, \citenamefont
  {Butch},\ and\ \citenamefont {Anlage}}]{Bae2021}%
  \BibitemOpen
  \bibfield  {author} {\bibinfo {author} {\bibfnamefont {S.}~\bibnamefont
  {Bae}}, \bibinfo {author} {\bibfnamefont {H.}~\bibnamefont {Kim}}, \bibinfo
  {author} {\bibfnamefont {Y.~S.}\ \bibnamefont {Eo}}, \bibinfo {author}
  {\bibfnamefont {S.}~\bibnamefont {Ran}}, \bibinfo {author} {\bibfnamefont
  {I.-l.}\ \bibnamefont {Liu}}, \bibinfo {author} {\bibfnamefont {W.~T.}\
  \bibnamefont {Fuhrman}}, \bibinfo {author} {\bibfnamefont {J.}~\bibnamefont
  {Paglione}}, \bibinfo {author} {\bibfnamefont {N.~P.}\ \bibnamefont
  {Butch}},\ and\ \bibinfo {author} {\bibfnamefont {S.~M.}\ \bibnamefont
  {Anlage}},\ }\bibfield  {title} {\bibinfo {title} {{Anomalous normal fluid
  response in a chiral superconductor UTe$_2$}},\ }\href
  {https://doi.org/10.1038/s41467-021-22906-6} {\bibfield  {journal} {\bibinfo
  {journal} {Nature communications}\ }\textbf {\bibinfo {volume} {12}},\
  \bibinfo {pages} {2644} (\bibinfo {year} {2021})}\BibitemShut {NoStop}%
\bibitem [{\citenamefont {Ishihara}\ \emph {et~al.}(2023)\citenamefont
  {Ishihara}, \citenamefont {Roppongi}, \citenamefont {Kobayashi},
  \citenamefont {Imamura}, \citenamefont {Mizukami}, \citenamefont {Sakai},
  \citenamefont {Opletal}, \citenamefont {Tokiwa}, \citenamefont {Haga},
  \citenamefont {Hashimoto} \emph {et~al.}}]{Ishihara2023}%
  \BibitemOpen
  \bibfield  {author} {\bibinfo {author} {\bibfnamefont {K.}~\bibnamefont
  {Ishihara}}, \bibinfo {author} {\bibfnamefont {M.}~\bibnamefont {Roppongi}},
  \bibinfo {author} {\bibfnamefont {M.}~\bibnamefont {Kobayashi}}, \bibinfo
  {author} {\bibfnamefont {K.}~\bibnamefont {Imamura}}, \bibinfo {author}
  {\bibfnamefont {Y.}~\bibnamefont {Mizukami}}, \bibinfo {author}
  {\bibfnamefont {H.}~\bibnamefont {Sakai}}, \bibinfo {author} {\bibfnamefont
  {P.}~\bibnamefont {Opletal}}, \bibinfo {author} {\bibfnamefont
  {Y.}~\bibnamefont {Tokiwa}}, \bibinfo {author} {\bibfnamefont
  {Y.}~\bibnamefont {Haga}}, \bibinfo {author} {\bibfnamefont {K.}~\bibnamefont
  {Hashimoto}}, \emph {et~al.},\ }\bibfield  {title} {\bibinfo {title} {{Chiral
  superconductivity in UTe$_2$ probed by anisotropic low-energy excitations}},\
  }\href {https://doi.org/10.1038/s41467-023-38688-y} {\bibfield  {journal}
  {\bibinfo  {journal} {Nature Communications}\ }\textbf {\bibinfo {volume}
  {14}},\ \bibinfo {pages} {2966} (\bibinfo {year} {2023})}\BibitemShut
  {NoStop}%
\bibitem [{\citenamefont {Jiao}\ \emph {et~al.}(2020)\citenamefont {Jiao},
  \citenamefont {Howard}, \citenamefont {Ran}, \citenamefont {Wang},
  \citenamefont {Rodriguez}, \citenamefont {Sigrist}, \citenamefont {Wang},
  \citenamefont {Butch},\ and\ \citenamefont {Madhavan}}]{Jiao2020}%
  \BibitemOpen
  \bibfield  {author} {\bibinfo {author} {\bibfnamefont {L.}~\bibnamefont
  {Jiao}}, \bibinfo {author} {\bibfnamefont {S.}~\bibnamefont {Howard}},
  \bibinfo {author} {\bibfnamefont {S.}~\bibnamefont {Ran}}, \bibinfo {author}
  {\bibfnamefont {Z.}~\bibnamefont {Wang}}, \bibinfo {author} {\bibfnamefont
  {J.~O.}\ \bibnamefont {Rodriguez}}, \bibinfo {author} {\bibfnamefont
  {M.}~\bibnamefont {Sigrist}}, \bibinfo {author} {\bibfnamefont
  {Z.}~\bibnamefont {Wang}}, \bibinfo {author} {\bibfnamefont {N.~P.}\
  \bibnamefont {Butch}},\ and\ \bibinfo {author} {\bibfnamefont
  {V.}~\bibnamefont {Madhavan}},\ }\bibfield  {title} {\bibinfo {title}
  {{Chiral superconductivity in heavy-fermion metal UTe$_2$}},\ }\href
  {https://doi.org/10.1038/s41586-020-2122-2} {\bibfield  {journal} {\bibinfo
  {journal} {Nature}\ }\textbf {\bibinfo {volume} {579}},\ \bibinfo {pages}
  {523} (\bibinfo {year} {2020})}\BibitemShut {NoStop}%
\bibitem [{\citenamefont {Thomas}\ \emph {et~al.}(2020)\citenamefont {Thomas},
  \citenamefont {Santos}, \citenamefont {Christensen}, \citenamefont {Asaba},
  \citenamefont {Ronning}, \citenamefont {Thompson}, \citenamefont {Bauer},
  \citenamefont {Fernandes}, \citenamefont {Fabbris},\ and\ \citenamefont
  {Rosa}}]{Thomas2020}%
  \BibitemOpen
  \bibfield  {author} {\bibinfo {author} {\bibfnamefont {S.}~\bibnamefont
  {Thomas}}, \bibinfo {author} {\bibfnamefont {F.}~\bibnamefont {Santos}},
  \bibinfo {author} {\bibfnamefont {M.}~\bibnamefont {Christensen}}, \bibinfo
  {author} {\bibfnamefont {T.}~\bibnamefont {Asaba}}, \bibinfo {author}
  {\bibfnamefont {F.}~\bibnamefont {Ronning}}, \bibinfo {author} {\bibfnamefont
  {J.}~\bibnamefont {Thompson}}, \bibinfo {author} {\bibfnamefont
  {E.}~\bibnamefont {Bauer}}, \bibinfo {author} {\bibfnamefont
  {R.}~\bibnamefont {Fernandes}}, \bibinfo {author} {\bibfnamefont
  {G.}~\bibnamefont {Fabbris}},\ and\ \bibinfo {author} {\bibfnamefont
  {P.}~\bibnamefont {Rosa}},\ }\bibfield  {title} {\bibinfo {title} {{Evidence
  for a pressure-induced antiferromagnetic quantum critical point in
  intermediate-valence UTe$_2$}},\ }\href
  {https://doi.org/10.1126/sciadv.abc8709} {\bibfield  {journal} {\bibinfo
  {journal} {Science Advances}\ }\textbf {\bibinfo {volume} {6}},\ \bibinfo
  {pages} {eabc8709} (\bibinfo {year} {2020})}\BibitemShut {NoStop}%
\bibitem [{\citenamefont {Thomas}\ \emph {et~al.}(2021)\citenamefont {Thomas},
  \citenamefont {Stevens}, \citenamefont {Santos}, \citenamefont {Fender},
  \citenamefont {Bauer}, \citenamefont {Ronning}, \citenamefont {Thompson},
  \citenamefont {Huxley},\ and\ \citenamefont {Rosa}}]{Thomas2021}%
  \BibitemOpen
  \bibfield  {author} {\bibinfo {author} {\bibfnamefont {S.~M.}\ \bibnamefont
  {Thomas}}, \bibinfo {author} {\bibfnamefont {C.}~\bibnamefont {Stevens}},
  \bibinfo {author} {\bibfnamefont {F.~B.}\ \bibnamefont {Santos}}, \bibinfo
  {author} {\bibfnamefont {S.~S.}\ \bibnamefont {Fender}}, \bibinfo {author}
  {\bibfnamefont {E.~D.}\ \bibnamefont {Bauer}}, \bibinfo {author}
  {\bibfnamefont {F.}~\bibnamefont {Ronning}}, \bibinfo {author} {\bibfnamefont
  {J.~D.}\ \bibnamefont {Thompson}}, \bibinfo {author} {\bibfnamefont
  {A.}~\bibnamefont {Huxley}},\ and\ \bibinfo {author} {\bibfnamefont
  {P.~F.~S.}\ \bibnamefont {Rosa}},\ }\bibfield  {title} {\bibinfo {title}
  {{Spatially inhomogeneous superconductivity in ${\mathrm{UTe}}_{2}$}},\
  }\href {https://doi.org/10.1103/PhysRevB.104.224501} {\bibfield  {journal}
  {\bibinfo  {journal} {Phys. Rev. B}\ }\textbf {\bibinfo {volume} {104}},\
  \bibinfo {pages} {224501} (\bibinfo {year} {2021})}\BibitemShut {NoStop}%
\bibitem [{\citenamefont {Choi}\ \emph {et~al.}(2018)\citenamefont {Choi},
  \citenamefont {Klein}, \citenamefont {Rosch},\ and\ \citenamefont
  {Kim}}]{Rosch2018}%
  \BibitemOpen
  \bibfield  {author} {\bibinfo {author} {\bibfnamefont {W.}~\bibnamefont
  {Choi}}, \bibinfo {author} {\bibfnamefont {P.~W.}\ \bibnamefont {Klein}},
  \bibinfo {author} {\bibfnamefont {A.}~\bibnamefont {Rosch}},\ and\ \bibinfo
  {author} {\bibfnamefont {Y.~B.}\ \bibnamefont {Kim}},\ }\bibfield  {title}
  {\bibinfo {title} {{Topological superconductivity in the Kondo-Kitaev
  model}},\ }\href {https://doi.org/10.1103/PhysRevB.98.155123} {\bibfield
  {journal} {\bibinfo  {journal} {Phys. Rev. B}\ }\textbf {\bibinfo {volume}
  {98}},\ \bibinfo {pages} {155123} (\bibinfo {year} {2018})}\BibitemShut
  {NoStop}%
\bibitem [{\citenamefont {Coleman}\ \emph {et~al.}(2022)\citenamefont
  {Coleman}, \citenamefont {Panigrahi},\ and\ \citenamefont
  {Tsvelik}}]{Coleman2022}%
  \BibitemOpen
  \bibfield  {author} {\bibinfo {author} {\bibfnamefont {P.}~\bibnamefont
  {Coleman}}, \bibinfo {author} {\bibfnamefont {A.}~\bibnamefont {Panigrahi}},\
  and\ \bibinfo {author} {\bibfnamefont {A.}~\bibnamefont {Tsvelik}},\
  }\bibfield  {title} {\bibinfo {title} {{Solvable 3D Kondo Lattice Exhibiting
  Pair Density Wave, Odd-Frequency Pairing, and Order Fractionalization}},\
  }\href {https://doi.org/10.1103/PhysRevLett.129.177601} {\bibfield  {journal}
  {\bibinfo  {journal} {Phys. Rev. Lett.}\ }\textbf {\bibinfo {volume} {129}},\
  \bibinfo {pages} {177601} (\bibinfo {year} {2022})}\BibitemShut {NoStop}%
\bibitem [{\citenamefont {Tsvelik}\ and\ \citenamefont
  {Coleman}(2022)}]{Tsvelik2022}%
  \BibitemOpen
  \bibfield  {author} {\bibinfo {author} {\bibfnamefont {A.~M.}\ \bibnamefont
  {Tsvelik}}\ and\ \bibinfo {author} {\bibfnamefont {P.}~\bibnamefont
  {Coleman}},\ }\bibfield  {title} {\bibinfo {title} {{Order fractionalization
  in a Kitaev-Kondo model}},\ }\href
  {https://doi.org/10.1103/PhysRevB.106.125144} {\bibfield  {journal} {\bibinfo
   {journal} {Phys. Rev. B}\ }\textbf {\bibinfo {volume} {106}},\ \bibinfo
  {pages} {125144} (\bibinfo {year} {2022})}\BibitemShut {NoStop}%
\bibitem [{\citenamefont {Yao}\ and\ \citenamefont {Lee}(2011)}]{Yao2011}%
  \BibitemOpen
  \bibfield  {author} {\bibinfo {author} {\bibfnamefont {H.}~\bibnamefont
  {Yao}}\ and\ \bibinfo {author} {\bibfnamefont {D.-H.}\ \bibnamefont {Lee}},\
  }\bibfield  {title} {\bibinfo {title} {{Fermionic Magnons, Non-Abelian
  Spinons, and the Spin Quantum {Hall} Effect from an Exactly Solvable
  Spin-$1/2$ {Kitaev} Model with {SU(2)} Symmetry}},\ }\href
  {https://doi.org/10.1103/PhysRevLett.107.087205} {\bibfield  {journal}
  {\bibinfo  {journal} {Phys. Rev. Lett.}\ }\textbf {\bibinfo {volume} {107}},\
  \bibinfo {pages} {087205} (\bibinfo {year} {2011})}\BibitemShut {NoStop}%
\bibitem [{\citenamefont {Coleman}\ \emph {et~al.}(1993)\citenamefont
  {Coleman}, \citenamefont {Miranda},\ and\ \citenamefont
  {Tsvelik}}]{Coleman1993}%
  \BibitemOpen
  \bibfield  {author} {\bibinfo {author} {\bibfnamefont {P.}~\bibnamefont
  {Coleman}}, \bibinfo {author} {\bibfnamefont {E.}~\bibnamefont {Miranda}},\
  and\ \bibinfo {author} {\bibfnamefont {A.}~\bibnamefont {Tsvelik}},\
  }\bibfield  {title} {\bibinfo {title} {Possible realization of odd-frequency
  pairing in heavy fermion compounds},\ }\href
  {https://doi.org/10.1103/PhysRevLett.70.2960} {\bibfield  {journal} {\bibinfo
   {journal} {Phys. Rev. Lett.}\ }\textbf {\bibinfo {volume} {70}},\ \bibinfo
  {pages} {2960} (\bibinfo {year} {1993})}\BibitemShut {NoStop}%
\bibitem [{\citenamefont {Kitaev}(2006)}]{Kitaev2006}%
  \BibitemOpen
  \bibfield  {author} {\bibinfo {author} {\bibfnamefont {A.}~\bibnamefont
  {Kitaev}},\ }\bibfield  {title} {\bibinfo {title} {Anyons in an exactly
  solved model and beyond},\ }\href
  {https://doi.org/https://doi.org/10.1016/j.aop.2005.10.005} {\bibfield
  {journal} {\bibinfo  {journal} {Annals of Physics}\ }\textbf {\bibinfo
  {volume} {321}},\ \bibinfo {pages} {2 } (\bibinfo {year} {2006})}\BibitemShut
  {NoStop}%
\bibitem [{\citenamefont {Lieb}(1994)}]{Lieb1994}%
  \BibitemOpen
  \bibfield  {author} {\bibinfo {author} {\bibfnamefont {E.~H.}\ \bibnamefont
  {Lieb}},\ }\bibfield  {title} {\bibinfo {title} {Flux phase of the
  half-filled band},\ }\href {https://doi.org/10.1103/PhysRevLett.73.2158}
  {\bibfield  {journal} {\bibinfo  {journal} {Phys. Rev. Lett.}\ }\textbf
  {\bibinfo {volume} {73}},\ \bibinfo {pages} {2158} (\bibinfo {year}
  {1994})}\BibitemShut {NoStop}%
\bibitem [{\citenamefont {Haldane}(1988)}]{Haldane1988}%
  \BibitemOpen
  \bibfield  {author} {\bibinfo {author} {\bibfnamefont {F.~D.~M.}\
  \bibnamefont {Haldane}},\ }\bibfield  {title} {\bibinfo {title} {{Model for a
  Quantum Hall Effect without Landau Levels: Condensed-Matter Realization of
  the "Parity Anomaly"}},\ }\href {https://doi.org/10.1103/PhysRevLett.61.2015}
  {\bibfield  {journal} {\bibinfo  {journal} {Phys. Rev. Lett.}\ }\textbf
  {\bibinfo {volume} {61}},\ \bibinfo {pages} {2015} (\bibinfo {year}
  {1988})}\BibitemShut {NoStop}%
\bibitem [{\citenamefont {Altland}\ and\ \citenamefont
  {Zirnbauer}(1997)}]{Altland1997}%
  \BibitemOpen
  \bibfield  {author} {\bibinfo {author} {\bibfnamefont {A.}~\bibnamefont
  {Altland}}\ and\ \bibinfo {author} {\bibfnamefont {M.~R.}\ \bibnamefont
  {Zirnbauer}},\ }\bibfield  {title} {\bibinfo {title} {Nonstandard symmetry
  classes in mesoscopic normal-superconducting hybrid structures},\ }\href
  {https://doi.org/10.1103/PhysRevB.55.1142} {\bibfield  {journal} {\bibinfo
  {journal} {Phys. Rev. B}\ }\textbf {\bibinfo {volume} {55}},\ \bibinfo
  {pages} {1142} (\bibinfo {year} {1997})}\BibitemShut {NoStop}%
\bibitem [{\citenamefont {Ludwig}(2015)}]{Ludwig2016}%
  \BibitemOpen
  \bibfield  {author} {\bibinfo {author} {\bibfnamefont {A.~W.~W.}\
  \bibnamefont {Ludwig}},\ }\bibfield  {title} {\bibinfo {title} {{Topological
  phases: classification of topological insulators and superconductors of
  non-interacting fermions, and beyond}},\ }\href
  {https://doi.org/10.1088/0031-8949/2015/T168/014001} {\bibfield  {journal}
  {\bibinfo  {journal} {Physica Scripta}\ }\textbf {\bibinfo {volume} {2016}},\
  \bibinfo {pages} {014001} (\bibinfo {year} {2015})}\BibitemShut {NoStop}%
\bibitem [{\citenamefont {Panigrahi}\ \emph {et~al.}(2024)\citenamefont
  {Panigrahi}, \citenamefont {Tsvelik},\ and\ \citenamefont
  {Coleman}}]{aaditya24}%
  \BibitemOpen
  \bibfield  {author} {\bibinfo {author} {\bibfnamefont {A.}~\bibnamefont
  {Panigrahi}}, \bibinfo {author} {\bibfnamefont {A.}~\bibnamefont {Tsvelik}},\
  and\ \bibinfo {author} {\bibfnamefont {P.}~\bibnamefont {Coleman}},\
  }\bibfield  {title} {\bibinfo {title} {{Breakdown of order fractionalization
  in the CPT model}},\ }\href {https://doi.org/10.1103/PhysRevB.110.104520}
  {\bibfield  {journal} {\bibinfo  {journal} {Phys. Rev. B}\ }\textbf {\bibinfo
  {volume} {110}},\ \bibinfo {pages} {104520} (\bibinfo {year}
  {2024})}\BibitemShut {NoStop}%
\end{thebibliography}%
\end{document}